\documentclass[a4paper,UKenglish,cleveref, autoref, nameinlink,thm-restate]{lipics-v2021}

\crefname{figure}{Figure}{Figures}
\Crefname{figure}{Figures}{Figures} 

\newlength{\foursep}
\newlength{\threesep}
\newlength{\twosep}
\usepackage{tikz}
\usepackage{wrapfig}
\usepackage{amssymb}
\usepackage{hyperref}
\usepackage{todonotes}
\usepackage{xcolor, soul}

\definecolor{defblue}{rgb}{0.121,0.47,0.705}
\definecolor{orangered}{rgb}{1,0.271,0}
\DeclareTextFontCommand{\emph}{\color{defblue}\em}

\definecolor{lipicsblue}{rgb}{0.08235294118,0.3098039216,0.537254902}

\hypersetup{colorlinks=true,
    linkcolor=lipicsblue,
    anchorcolor=lipicsblue,
    citecolor=lipicsblue,
    filecolor=lipicsblue,
    menucolor=lipicsblue,
    urlcolor=lipicsblue,
    bookmarksopen=true,
    bookmarksopenlevel=2,
    bookmarksnumbered=true,
    plainpages=false,
    }

\pdfoutput=1 
\hideLIPIcs  

\title{High-Modularity Graph Partitioning Through NLP Techniques and Maximal Clique Enumeration} 

\titlerunning{Graph Partitioning Through NLP Techniques and Maximal Clique Enumeration} 

\author{Marco {D'Elia}}{Roma Tre University, Rome, Italy}{marco.delia@uniroma3.it}{https://orcid.org/0009-0008-6266-3324}{}

\author{Irene Finocchi}{Luiss Guido Carli, Rome, Italy}
{finocchi@luiss.it}{https://0000-0002-6394-6798}{}

\author{Maurizio Patrignani}{Roma Tre University, Rome, Italy}{maurizio.patrignani@uniroma3.it}{https://orcid.org/0000-0001-9806-7411}{}

\authorrunning{M. D'Elia et al.}

\Copyright{Marco D'Elia, Irene Finocchi, and Maurizio Patrignani} 



\category{} 

\relatedversion{} 

\funding{This research was supported in part by MUR PRIN Projects EXPAND and NextGRAAL [grant numbers 2022TS4Y3N and 2022ME9Z78].}

\nolinenumbers 

\EventEditors{John Q. Open and Joan R. Access}
\EventNoEds{2}
\EventLongTitle{42nd Conference on Very Important Topics (CVIT 2016)}
\EventShortTitle{CVIT 2016}
\EventAcronym{CVIT}
\EventYear{2016}
\EventDate{December 24--27, 2016}
\EventLocation{Little Whinging, United Kingdom}
\EventLogo{}
\SeriesVolume{42}
\ArticleNo{23}

\begin{document}

\maketitle

\begin{abstract}
Natural Language Processing (NLP) provides highly effective tools for interpreting and handling human language, offering a broad spectrum of applications.
In this paper, we address a classic combinatorial problem -- finding graph partitions with high modularity -- by applying NLP techniques that compute term frequency and inverse document frequency (TF-IDF) alongside machine learning clustering algorithms.
We present a new framework, called {\em Clique-TF-IDF}, designed for graph partitioning, a task that holds significant relevance across various network analysis contexts. This approach uses dense substructures of the graph, specifically maximal cliques, to represent each vertex in terms of the cliques it is part of, in a manner akin to term-document matrices.
Experiments show that {\em Clique-TF-IDF} yields results that are comparable to or outperform the current state-of-the-art algorithms, whether or not the number of partitions is known in advance.
Although this framework emphasizes on cliques and partitioning, it can be extended to devise AI-driven solutions for a variety of challenging combinatorial problems that can leverage efficiently enumerable substructures.
\end{abstract}

\section{Introduction}

Community detection is a fundamental tool in network analytics, with widespread applications across various domains, including social networks, biology and bioinformatics, Internet and Web, recommendation systems, cybersecurity, transportation and infrastructure networks, financial networks, neuroscience, and criminal networks.

Intuitively, a community is a subset of vertices that are tightly connected among them. When every possible pair of vertices are adjacent, the community is fully connected and can be modeled by a {\em clique}. 
Graph-theoretic properties based on cliques, such as $k$-plexes, $s$-defective, or $\gamma$-quasi cliques, yield the strictest definitions of community. We refer to~\cite{Pattillo2012} for a survey. 

Although computing maximal cliques is computationally hard~\cite{DBLP:conf/coco/Karp72}, several centralized~\cite{DBLP:journals/cacm/BronK73,DBLP:journals/tcs/CazalsK08,DBLP:conf/wea/EppsteinS11,DBLP:journals/tcs/Koch01,DBLP:conf/cocoon/TomitaTT04} and distributed~\cite{DBLP:journals/tods/ChengKFYZ11,DBLP:conf/kdd/ChengZKC12,DBLP:conf/edbt/ConteVMPT16,xu2014distributed} algorithms have been developed to efficiently enumerate them in real-world networks.
However, the number of graph-theoretic communities is usually huge and their overlap is too large to allow the user to exploit such communities to analyze and decompose the network.
Summarization, compression, and hierarchical clustering methods have been proposed to overcome some of these limitations~\cite{DBLP:journals/csr/DEliaFP25,DBLP:journals/isci/GlariaHLNS21,DBLP:journals/isci/DrazdilovaPPS24}.

In contrast to graph-theoretic approaches, a more intuitive definition of community for users is that of partitioning, where the vertex set is divided into distinct disjoint blocks.   
Measures such as modularity capture the quality of the partition (we refer to \cref{se:background} for a formal definition). Unfortunately, computing partitions with maximum modularity is NP-complete~\cite{DBLP:conf/wg/BrandesDGGHNW07} and exact algorithms are not feasible, even for sparse networks. As a consequence, most existing approaches to network partitioning rely on heuristics (see \cref{se:related} for an overview on the state of the art). 

\smallskip
\noindent{\bf Contributions of the paper.} 
In this paper, we aim to establish a connection between communities defined in a graph-theoretic sense and vertex partitions. While the former can be computed using exact algorithms, the calculation of the latter relies on heuristic methods.

By using graph-theoretic communities as a guide, our approach applies an AI-driven method to the vertex partitioning problem.
In more detail, we begin by encoding maximal cliques into a matrix, with rows representing vertices and columns corresponding to cliques, akin to term-document matrices used in NLP domain. Through a weighting operation performed on this matrix, similar to function TF-IDF, we then assign higher values to  maximal cliques that are large and loosely connected to the rest of the network.

This matrix, once normalized, can be viewed as an embedding of the graph vertices into a metric space.
Finally, we cluster the rows of the matrix (i.e., the points in the metric space) according to their similarity (i.e., distance) in order to partition the network. 

This approach proves beneficial both when the number $k$ of blocks of the partition is specified in advance, and when such a number is not specified. For the first scenario, we produced two algorithms, dubbed $k$-{\em Aggl-Clique-TF-IDF} and $k$-{\em Means-Clique-TF-IDF}, that use an agglomerative approach~\cite{10.5555/3208440} or the $k$-Means algorithm~\cite{kmeans} to cluster the matrix rows, respectively. In \cref{sec:experiments-fixed-k} we experimentally show that the two algorithms outperform the well-known METIS graph partitioning algorithm~\cite{karypis1998fast} in terms of quality of the produced partitions,  measured both with respect to absolute partitioning metrics and with respect to the ground truth, at the cost of an increased computation time.

When the number $k$ of blocks of the partition is not provided in advance, building upon preliminary results presented in~\cite{d2023clique}, we add a further step to the approach in order to find a good value for $k$. The obtained algorithm, named {\em Clique-TF-IDF}, is experimentally compared with several state-of-the-art approaches in the literature considering both real-world and synthetic networks. The experiments reported in \cref{se:experiments} show that {\em Clique-TF-IDF} is among the approaches with the best performance, although the computation times are longer than those of Leiden~\cite{Leiden2019}, which is a good algorithm both with respect to effectiveness and with respect to efficiency. 

In order to validate the use of maximal cliques of a graph to obtain an embedding of its vertices into a metric space for clustering purposes, which is proposed for the first time in this paper, we also performed experiments where this embedding phase is replaced by alternative state-of-the-art graph embedding approaches. \cref{sec:experiments-embedding-phase} shows that the adopted embedding phase is more effective, more efficient, and also more practical, as it does not require many tuning parameters as is the case for alternative approaches.

As an additional contribution, in \cref{sse:setup} we also show how to generate a benchmark of networks whose ground truth has the same modularity independently of their size. 
This result can be of interest in its own right.

While our focus in this paper is on maximal cliques and partitioning algorithms, we believe that the strategy proposed  in this paper has broader potential and could be extended to develop AI-based solutions for various intractable combinatorial problems. These are problems where certain substructures can be efficiently enumerated and leveraged. 
For instance, listing independent sets might prove useful to color the vertices of a graph, since an independent set corresponds to a maximal clique in the complement graph. Similarly, max-cut problems might benefit from the efficient computation of maximal bipartite subgraphs~\cite{conte-bipartite}.
By combining combinatorial methods with AI techniques, these hybrid approaches offer the best of both domains. Moreover, they have the potential to open new application areas for multidisciplinary research opportunities.

\smallskip
\noindent{\bf Organization of the paper.} After surveying related work in \cref{se:related}, in \cref{se:background} we introduce notation and provide preliminary background.
Our approach is described in detail in \cref{se:methodology}.
The experimental setup is provided in \cref{sse:setup}, while
\cref{sec:experiments-fixed-k,se:experiments,sec:experiments-embedding-phase} discuss the experimental analyses. {\em Clique-TF-IDF} is compared to state-of-the-art methods in terms of both efficiency and effectiveness, focusing on different scenarios: when the number of communities is known in advance, when it is not known, and when only the embedding phase is considered.
Conclusions and directions for future work are outlined in \cref{se:conclusions}.

\section{Related work}
\label{se:related}
The graph partitioning problem has attracted significant interest from  both theoretical and practical perspectives and has been extensively studied. Various problem variants have been considered in the literature, with some works employing AI techniques to tackle different issues.
For instance, when vertices have associated attributes that must be considered in the partitioning, contrastive learning and graph neural networks facilitate the learning of clusterable features, resulting in partitions that align well with the ground truth~\cite{DBLP:conf/nips/DevvritSD022}. Algorithms extending traditional approaches for partitioning attributed graphs are discussed in~\cite{DBLP:conf/ida/CombeLGE15}.

In correlation partitioning, the edges are weighted quantitatively, representing vertex similarity. In this case, the objective is to maximize a measure that typically increases when edges with negative weights span across different partitions. This problem has been studied in~\cite{DBLP:journals/ml/BansalBC04,DBLP:conf/esa/SahaS19}, and an online version has been studied in~\cite{DBLP:conf/nips/LattanziMVWZ21}.

Another variant is structural clustering, where vertices in the same partition blocks are expected to have similar ``roles'' within the graph. In this context, graph neural networks have been applied to vertex classification and graph classification tasks~\cite{DBLP:journals/nn/ChenYHLPWZ23,DBLP:conf/cikm/CuiL0Y22,DBLP:conf/sdm/ZhuLHK21}. Additionally, structural vertex embedding techniques can be applied to identify an effective partition~\cite{DBLP:conf/kdd/RibeiroSF17}.

Differently from the aforementioned works, the focus of this paper is on positional clustering for graphs that have no attributes: vertex similarity depends on adjacency, and the partition quality is evaluated using modularity. This problem has been addressed in the literature in many previous works that present a variety of combinatorial algorithms, both when the number $k$ of desired partition blocks is known in advance and when it is not. For the former case, the well-known METIS approach~\cite{karypis1998fast} can be applied in either its recursive bisection or multiway partitioning variants. A non-negative matrix factorization-based algorithm is detailed in~\cite{al2023community}.

The case where $k$ is not part of the input has attracted more attention in the literature. Agglomerative hierarchical methods that maximize modularity measure are discussed in~\cite{PhysRevE.70.066111,Blondel_2008,Leiden2019}. Other approaches combine random walks on the graph with a similarity matrix~\cite{https://doi.org/10.48550/arxiv.physics/0512106} or use information theory methods~\cite{doi:10.1073/pnas.0706851105}. Label propagation techniques are considered in~\cite{Cordasco2010CommunityDV}, while statistical physics models are used in~\cite{Reichardt_2006}. Additional heuristic approaches, which generally show lower effectiveness, are discussed in~\cite{Girvan_2002,10.5555/2951659.2951772,10.1145/2566486.2568010}.

Finally, vertex embeddings based on positional information can also be leveraged to compute a partitioning~\cite{DBLP:conf/kdd/GroverL16,DBLP:conf/kdd/PerozziAS14,DBLP:conf/sdm/ZhuLHK21}, like in the case of structural clustering. For a comprehensive survey on vertex embedding techniques, particularly those leveraging graph neural network models, see~\cite{DBLP:journals/aiopen/LiuT21}. When vertices have associated attributes, embedding methods are further detailed in~\cite{DBLP:journals/aiopen/ZhaoCCZT22}. 

\section{Background}\label{se:background}
For an undirected graph $G$, we let $V(G)$ and $E(G)$ be its vertex and edge sets, respectively. Denote the number of vertices and edges of $G$ by $n$ and $m$, respectively. Let $A$ be the adjacency matrix of $G$, where $A_{uv} = 1$ if vertex $u$ and vertex $v$ are adjacent, and $0$ otherwise. In an undirected graph, $A$ is symmetric, so $A_{uv} = A_{vu}$. The degree of a vertex $u$ is denoted by $\delta(u)$. A \emph{partition of $G$} is a subdivision of $V$ into distinct blocks $\{V_1, \ldots, V_k\}$, such that $\bigcup_{i=1}^{k} V_i = V$ and $V_i \cap V_j = \varnothing$ for every $i \neq j$. Each block is intended to represent an internally dense area that is weakly connected to the other vertices.

\medskip \noindent{\bf Modularity.} 
\emph{Modularity} can be used to measure  the quality of graph partitions and is defined in~\cite{PhysRevE.70.066111} as follows:
\begin{equation}
	Q = \frac{1}{2m} \sum_{i=1}^k\sum_{u,v \in V_i} \left( A_{uv} - \frac{\delta_u \delta_v}{2m}\right)
\label{formula:modularity}
\end{equation}
Here $\frac{\delta_u \delta_v}{2m}$ 
denotes the probability that an edge exists between vertices 
$u$ and $v$ in a random network model that maintains the vertices' degree distribution.
The modularity measure $Q$ ranges in the interval $\left[-\frac{1}{2},1\right]$, where non-zero values indicate deviations from randomness, with higher values suggesting a stronger community structure.

In spite of the fact that maximizing modularity is NP-hard~\cite{DBLP:conf/wg/BrandesDGGHNW07}, this measure is widely used in previous works to assess the quality of communities, particularly in cases where a ground truth is unavailable or requires validation~\cite{Blondel_2008,PhysRevE.70.066111,Leiden2019}.

\medskip \noindent{\bf Permanence.} This is a more recent quality measure for graph partitions introduced in~\cite{chakraborty2014permanence}. We first need some preliminary notation:
\begin{itemize}
    \item $b(v)$ is the block of the partition $v$ belongs to;
    \item $I(v)$ is the number of edges incident to $v$ and internal to $b(v)$;
    \item $E_{max}(v)$ is the maximum number of edges of $v$ to a single block of the partition different from $b(v)$;
    \item $c_{in}(v)$ is the internal clustering coefficient of $b(v)$, i.e., the ratio of the existing edges and the total number of possible edges among the neighbors of $v$ that are part of $b(v)$.  
\end{itemize}

\noindent \emph{Permanence} is defined as:
$P(G) = \frac{1}{n}\sum_{v \in V} P(v)$ where:
\begin{equation}
    P(v) = \left(\frac{I(v)}{\max\,\{1,E_{max}(v)\}} \cdot \frac{1}{\delta_v}\right) - (1 - c_{in}(v))
\end{equation}

\noindent Observe that $P(v) \in [-1,1]$. The maximum value of $P(v)$ is equal to $c_{in}(v)=1$ and is obtained when $b(v)$ is an isolated clique of $G$ (in this case $I(v) = \delta(v)$). On the other side, $P(v)$ is close to $-1$ when $I(v) \ll \delta(v)$ and $c_{in}(v) = 0$. Consequently, also $P(G) \in [-1,1]$.

\medskip \noindent{\bf TF-IDF.} 
The term frequency–inverse document frequency (in short, TF-IDF) measures the importance of terms in documents from a collection $D = \{d_1, \ldots, d_r\}$ of size $r$. We assume terms to be taken from a set $T = \{t_1, \ldots, t_s\}$ of size $s$, and denote by $\tau_{t,d}$ the number of occurrences of a term $t\in T$ within a document $d\in D$. Since some terms may be more frequent in general, we count in $\delta_t$ the number of documents containing $t$ and define the \emph{inverse document frequency} of $t$ as $\gamma_t = \log(\frac{|D|}{\delta_t})$.
TF-IDF is a weighting function defined as $\omega_{t,d} = \tau_{t,d} \cdot \gamma_t$~\cite{manning2008introduction}: its values are higher for terms appearing in a small subset of documents, and lower when a term is not common within a document or is common in the entire corpus.

\medskip\noindent{\bf Hierarchical clustering.} Given a set $P=\{p_1, \ldots, p_n\}$ of $n$ points within a $d$-dimensional space and an integer $k$, the objective of geometric clustering is to divide $P$ into $k$ non-empty clusters $\{P_1, \ldots, P_k\}$ such that $\bigcup_{i=1}^{k} P_i = P$ and $P_i \cap P_j = \varnothing$ for every $i \neq j$. A well-known solution is based on the $k$-Means algorithm~\cite{kmeans}.

Another well-known method to compute a geometric clustering of $P$ is hierarchical clustering, valued for its flexibility in identifying both local and global structures. One of the most popular variants is called {\em agglomerative clustering}~\cite{10.5555/3208440} and begins by treating each individual point as a separate cluster, successively merging them into larger clusters. This approach relies on three key components: (i) a metric for calculating distances between points, (ii) a ``linkage criterion'' to measure distances or similarities between clusters to determine which to merge, and (iii) a ``stopping criterion'' to end the computation at an appropriate hierarchical level.
Unlike other clustering algorithms based on density or centroids, hierarchical approaches do not require any initialization. Moreover, it benefits from only using distances, rather than absolute coordinates, making it more broadly applicable.

\section{Overview of the approach}\label{se:methodology}

This section introduces our graph partitioning method. We leverage the efficient enumeration of the maximal cliques of a graph $G$ and follow a pipeline consisting of the following macro-steps:

\begin{description}

\item{\textsc{Vertex-community matrix:}} after computing all the maximal cliques, we construct a vertex-community matrix~$Z$, with values reflecting the sizes of the cliques.

\item{\textsc{Tf-idf transformation:}} a weighting operation similar to TF-IDF is applied to~$Z$, assigning higher values to maximal cliques that are both large and weakly connected to the rest of the network. 

\item{\textsc{Clustering:}} by regarding the matrix rows as a set of points embedded in a multi-dimensional space, the graph partition is ultimately determined by clustering these points.

\end{description}

\subsection{Vertex-community matrix}

To enumerate maximal cliques, we use the algorithm introduced in~\cite{DBLP:journals/cacm/BronK73}. We denote by $C = \{c_1, \ldots, c_d\}$ the set of all maximal cliques. Let $n$ and $d$ be the number of vertices and the number of maximal cliques of $G$, respectively. We use two auxiliary data structures.

\smallskip \noindent{\bf Clique-incidence matrix Y.} Vertex containment in maximal cliques is described by a \emph{clique-incidence} matrix $Y \in \mathbb{N}^{n \times d}$. More formally:
$$
Y_{i\ell} = \left\{
\begin{array}{ll}
1 \hspace{1mm} & \mbox{if vertex $v_i$ belongs to maximal clique $c_\ell$}\\
0             & \mbox{otherwise}
\end{array}
\right.
$$

\begin{figure*}[t]
  \centering
  \begin{tabular}{  c  c  }
  \raisebox{-.5\height}{\includegraphics[width=2.0cm]{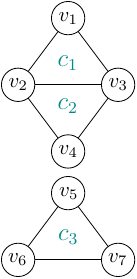}}
  &
  \small
  \begin{tabular}{|r||c|c|c|c|c|c|c|}
  \hline
    & $v_1$ & $v_2$ & $v_3$ & $v_4$ & $v_5$ & $v_6$ & $v_7$\\ \hline\hline
  $v_1$ & 3 & 3 & 3 & 0 & 0 & 0 & 0\\ \hline
  $v_2$ & 3 & 6 & 6 & 3 & 0 & 0 & 0\\ \hline
  $v_3$ & 3 & 6 & 6 & 3 & 0 & 0 & 0\\ \hline
  $v_4$ & 0 & 3 & 3 & 3 & 0 & 0 & 0\\ \hline 
  $v_5$ & 0 & 0 & 0 & 0 & 3 & 3 & 3\\ \hline
  $v_6$ & 0 & 0 & 0 & 0 & 3 & 3 & 3\\ \hline
  $v_7$ & 0 & 0 & 0 & 0 & 3 & 3 & 3\\ \hline
  \end{tabular}
  \\
  \begin{minipage}[t][1.5cm][c]{2.5cm}
    \begin{center}
      (a) Graph $G$         
    \end{center}
  \end{minipage}
  & 
  \begin{minipage}[t][1.5cm][c]{6cm}
    \begin{center}
      (b) Co-participation matrix $X$         
    \end{center}
  \end{minipage}
\\
  \small
  \begin{tabular}{|r || c | c | c |}
  \hline
    & \textcolor{teal}{$c_1$} & \textcolor{teal}{$c_2$} & \textcolor{teal}{$c_3$} \\ \hline\hline
  $v_1$ & 1 & 0 & 0 \\ \hline
  $v_2$ & 1 & 1 & 0 \\ \hline
  $v_3$ & 1 & 1 & 0 \\ \hline
  $v_4$ & 0 & 1 & 0 \\ \hline
  $v_5$ & 0 & 0 & 1 \\ \hline
  $v_6$ & 0 & 0 & 1 \\ \hline
  $v_7$ & 0 & 0 & 1 \\ \hline
  \end{tabular}
  &
  \small
  \begin{tabular}{|r || c | c | c |}
  \hline
    & \textcolor{teal}{$c_1$} & \textcolor{teal}{$c_2$} & \textcolor{teal}{$c_3$} \\ \hline\hline
  $v_1$ & 9 & 6 & 0 \\ \hline
  $v_2$ & 15 & 15 & 0 \\ \hline
  $v_3$ & 15 & 15 & 0 \\ \hline
  $v_4$ & 6 & 9 & 0 \\ \hline
  $v_5$ & 0 & 0 & 9 \\ \hline
  $v_6$ & 0 & 0 & 9 \\ \hline
  $v_7$ & 0 & 0 & 9 \\ \hline
  \end{tabular}
  \\ \\
  \begin{minipage}[t][0.5cm][c]{6.1cm}
     \begin{center}
      (c) Clique-incidence matrix $Y$         
     \end{center}
  \end{minipage}
  & 
  \begin{minipage}[t][0.5cm][c]{6.1cm}
     \begin{center}
      (d) Vertex-community matrix $Z$          
     \end{center}
  \end{minipage}
  \\
  \end{tabular}
  \caption{(a) An example graph $G$. The matrix $Z$ (d) is the product of $X$ (c) and $Y$ (b).} 
  \label{fi:toy-example}
\end{figure*}

\noindent As an example, the incidence matrix $Y$ depicted in \cref{fi:toy-example}(c) shows clique membership for the graph of \cref{fi:toy-example}(a), which contains three maximal cliques $c_1$, $c_2$, and $c_3$. Our goal is to cluster the rows of $Y$ based on their similarity, grouping in the same partition vertices belonging to very similar sets of maximal cliques. For instance, in \cref{fi:toy-example}(c), vertices  $v_1$ and $v_2$ share more similarities compared to vertices $v_1$ and $v_5$.

\smallskip \noindent{\bf Co-participation matrix X.} 
Directly using the matrix $Y$ without any post-processing would not emphasize whether the vertices belong to maximal cliques of large or small size.
Also, we would like that two vertices that do not fall into the same maximal clique (see, for example, vertices $v_1$ and $v_4$ of graph $G$ in \cref{fi:toy-example}(a)) had some similarity in matrix $Y$ based on the fact that they have neighbors in common in some other maximal clique (see, for example, the rows $1$ and $4$ in \cref{fi:toy-example}(b) corresponding to vertices $v_1$ and $v_4$).

In order to solve the above two problems, we introduce an auxiliary matrix  $X \in \mathbb{R}^{n \times n}$, called \emph{co-participation} matrix. Each entry $X_{ij}$ of matrix $X$ accounts for the sizes of the maximal cliques to which vertices $v_i$ and $v_j$ both belong. Precisely, let $w_\ell$ be the number of edges of a maximal clique $c_\ell$. For any pair of vertices $\langle v_i, v_j \rangle$, where $i,j = 1, \dots, n$ and $i\neq j$, we set $X_{ij} = \sum_{v_i, v_j \in c_\ell} w_\ell$. As for the diagonal of matrix $X$, we set $X_{ii} = \sum_{v_i \in c_\ell} w_\ell$. Observe that matrix $X$ is symmetric. For the example graph of \cref{fi:toy-example}(a), since both vertices $v_2$ and $v_3$ share the maximal clique $c_1$, with $w_1 = 3$, and the maximal clique $c_2$, with $w_2 = 3$, we have $X_{2\,3} = w_1 + w_2 = 6$.

\smallskip \noindent{\bf Vertex-community matrix Z.} 
We use matrix $X$ to process matrix $Y$ performing the matrix multiplication $Z = X \cdot Y$. We call the obtained matrix $Z$ \emph{vertex-community matrix}. Observe that $Z$ has the same size $\mathbb{R}^{n \times d}$ as matrix $Y$. 
Each entry $Z_{i\ell}$ of $Z$ can be viewed as the sum of two types of terms: 
\begin{itemize}

\item \emph{Actual involvement} of vertex $v_i$ into the clique $c_\ell$: when $v_i \in c_\ell$ we sum to $Z_{i\ell}$ the value $|c_\ell| \cdot w_\ell$ (see for example $Z_{4\,2}$ in \cref{fi:toy-example}(d), where $Z_{4\,2} = |c_2| \cdot w_2 = 3 \times 3 = 9$). This value quadratically increases with the number of vertices of $c_\ell$.

\item \emph{Transitive involvement} of vertex $v_i$ into the clique $c_\ell$: when $v_i \notin c_\ell$, for each vertex $v_j \in c_\ell$ we sum to $Z_{i\ell}$ the term $w_{\ell'}$, where $c_\ell'$ is a maximal clique to which both $v_i$ and $v_j$ belong. For example, vertex $v_4$ has a transitive involvement in clique $c_1$ because of its neighbors $v_2$ and $v_3$. Hence, value $Z_{4\,1} = 6$ is the sum of the two terms $X_{4\,2} \times Y_{2\,1} = w_2 = 3$ and $X_{4\,3} \times Y_{3\,1} = w_2 = 3$.

\end{itemize}
\subsection{TF-IDF transformation}

Matrix $Z$ can be compared to the term-frequency matrix used in Information Retrieval, where occurrences of terms in documents are encoded. In our case, vertices correspond to documents and maximal clique correspond to terms. Following this analogy, we aim to assign more importance to a maximal clique (term) for a vertex (document) when the corresponding column in $Z$ is sparse (i.e., the term is infrequent). This is because a maximal clique that is weakly connected to the rest of the graph is less desirable to split across different blocks of the partition.
We therefore compute a vector $\Gamma = [\gamma_1, \dots, \gamma_d] \in \mathbb{R}^{d}$, where, for each $\ell = 1, \dots, d$, $\gamma_\ell =\log({\frac{n}{|\delta_\ell|}})$, where $\delta_\ell$ is the number of non-zero elements of column $\ell$ of $Z$, analogous to the inverse document frequency. We then modify $Z$ by performing, for each $i = 1, \dots, n$, an element-wise multiplication of row $Z_i$ with $\Gamma$, i.e., $Z_i$ is replaced by $Z_i \odot \Gamma$, where $\odot$ denotes the Hadamard product between two vectors.
In the example of \cref{fi:toy-example}, after the transformation described above, the value for the element $Z_{1\,2}$ for clique $c_2$ with respect to vertex $v_1$ is $\omega_{1,2} = \tau_{1,2} \cdot \gamma_1 = Z_{1\,2} \cdot \log({\frac{n}{|\delta_2|}}) = 6 \cdot 0.8 = 4.8$. 

Finally, to obtain an embedding for the vertices, we normalize each row of $Z$ so that each vertex corresponds to a unit-length vector.

\subsection{Clustering}

Matrix $Z$ can be viewed as a set of $n$ points (the vertices in $V$) into a $d$-dimensional space (the cliques).
Hence, we can perform geometric clustering of the graph by partitioning these $n$ points into a specified number $k$ of clusters. To achieve this, we apply two alternative methods:
(i) an agglomerative clustering approach~\cite{10.5555/3208440}, which initially assigns each point to its own cluster and then iteratively merges the two closest clusters, where the distance between clusters is defined as the average distance between their vertices. This process continues until the desired number of clusters, $k$, is reached.
We call the obtained algorithm $k$-{\em Aggl-Clique-TF-IDF};
(ii) $k$-Means algorithm~\cite{kmeans}, a reknown iterative centroid-based approach to produce $k$-way partitions of points embedded into a metric space. We call the obtained algorithm $k$-{\em Means-Clique-TF-IDF}.

\smallskip \noindent{\bf Choosing the number of clusters.} 
Since the algorithm requires the number of clusters $k$ as input, we employ a straightforward heuristic by performing a binary search over possible $k$ values to find one that achieves a high modularity score.

In fact, we experimentally assessed that the modularity of the partition computed on matrix $Z$ is usually very close to a bitonic curve with respect to number $k$ of blocks of the partition. This empirical property ensures the effectiveness of the binary search. 
We consider only agglomerative clustering in this scenario. The efficiency of this search procedure depends indeed by two facts: first, the agglomerative clustering hierarchy is computed once at the start and reused to generate $k$-way partitions for different values of $k$; second, when computing the agglomerative hierarchy only the distances among points are used, instead of their actual coordinates. We call the obtained algorithm {\em Clique-TF-IDF}.


\section{Experimental setup}\label{sse:setup}

In this section we discuss quality metrics (\cref{ss:metrics}), dataset generation (\cref{ss:datasets}), and algorithms under evaluation (\cref{ss:algorithms}). All experiments were conducted on a laptop with an Intel\textregistered{} Core\texttrademark{} i5-1135G7 CPU (4 cores, 4.2 GHz) and 8 GB of RAM.  The selection of parameters in the generation of synthetic networks, as discussed in \cref{sss:synthetic}, is especially interesting and can be regarded as an additional contribution of the paper.

%
%
\subsection{Metrics}
\label{ss:metrics}
In order to compare the effectiveness of the partitioning algorithms, we contrast them with respect to the modularity and the permanence of the produced partitions, which are standard evaluation metrics in this domain. For the synthetic instances, where a ground truth is also available, we also measure the normalized mutual information (NMI)~\cite{manning2008introduction}.

%
%
\subsection{Datasets}\label{ss:datasets}

In our experiments we used two kinds of datasets: real-world networks and synthetic networks. Real-world networks offer a measure of the performances of the algorithms in application contexts. Synthetic networks allow us to explore in a more systematic way the space of the possible inputs. Further, they allow us to compare the output partitioning with the ground truth used to produce the input instances, yielding an additional quality measure.

%
%
\subsubsection{Real-world networks} 

We considered a dataset with ten real-world networks of varying sizes, ranging from tens to ten thousand vertices, collected from open repositories~\cite{snapnets,KONECT,networkrepository}, which are commonly used as benchmarks in this field. Their characteristics, including the number of maximal cliques, are listed in \cref{ta:dataset}. Entries marked with $^*$ correspond to the largest component of networks, as certain algorithms, such as Pott~\cite{Reichardt_2006}, require connected networks. These components usually include a significant portion of the vertices. Additionally, We pre-processed the networks by removing loops and multiple edges. Networks originally containing loops and/or multiple edges are marked with $^{**}$ in \cref{ta:dataset}. The values reported in \cref{ta:dataset} refer to the networks after the operations described above. 

\begin{table*}[htb]
\small
\centering
\caption{Real-world networks used in the experimental analysis. Entries labeled with $^*$ only refer to the giant component of the networks. Entries labeled with $^{**}$ refer to networks where loops and multiple edges have been removed.}\label{ta:dataset}
\begin{tabular}{|r||c|c|c|c|}
\hline

Network & Alias & vertices & Edges & \# Maximal cliques \\ \hline\hline
arenas-email & arenas & 1133 & 5451 & 3267 \\ \hline
arxiv-grqc$^*$ & grqc & 4158 & 13422 & 3385 \\ \hline
arxiv-hepth$^*$ & hepth & 8638 & 24806 & 9357 \\ \hline
arxiv-hepph$^*$ & hepph & 11204 & 117619 & 14588 \\ \hline
citeseer$^{**}$ & cite & 2120 & 3679 & 2722 \\ \hline
email-eu-core$^{**}$ & email & 986 & 16064 & 42709 \\ \hline
feather-lastfm-social & lastfm & 7624 & 27806 & 17957 \\ \hline
p2p-gnutella04 & p2p & 10876 & 39994 & 38497 \\ \hline
sociopatterns-infectious & socio & 410 & 2765 & 1247 \\ \hline
zacharys-karate & karate & 34 & 78 & 36 \\ \hline
\end{tabular}
\end{table*}

%
%
\subsubsection{Synthetic networks} 
\label{sss:synthetic}

We used the LFR~\cite{PhysRevE.78.046110} library to produce synthetic networks with {\em a priori} known communities where both the vertex degree and the community size distribution are power laws. This library is specifically aimed at comparing different methods for community detection and graph partitioning. 
Our purpose is that of producing a benchmark of networks to show how some parameters of the instances impact on the performances of the algorithms. In particular, we selected the following parameters:
\begin{itemize}
    \item Number $n$ of vertices, which is a standard parameter to measure the time complexity of network algorithms. 
    \item Average vertex degree $\langle d \rangle$: this is related to the number of edges in the network, which is a second standard parameter considered in network classification.
    \item Maximum vertex degree $d_{max}$: this is a measure of how much evenly the edges are distributed in the network. Scale-free networks have a maximum vertex degree much higher than the average vertex degree and often proportional to the number $n$ of vertices. 
    \item Noise $\mu$ (also referred to as ``mixing parameter''): this is related to the community structure and determines of how much the internal links of each community are rewired towards vertices of other communities. If this parameter were set to $1$, no links would survive within the community, contradicting our intuitive notion of community.
\end{itemize}

\noindent The LFR library guarantees a scale-free distribution by choosing the number of communities of the network and their sizes. 
By varying parameters $n$, $\langle d \rangle$, $d_{max}$, and $\mu$ linearly in their allowed intervals, several combinations do not correspond to feasible networks and the LFR library does not produce any instance. On the other hand, instances produced by the LFR library but having a low value of modularity with respect to the ground truth are not suitable for an experimental analysis of partitioning algorithms: the ground truth becomes challenging, if not impossible, to retrieve in such instances, primarily due to the overwhelming noise that obscures it. With the purpose of obtaining a benchmark of synthetic scale-free networks featuring significant communities we used the following values for the chosen parameters.

\begin{figure*}[tb]
\centering
\includegraphics[width=1\linewidth]{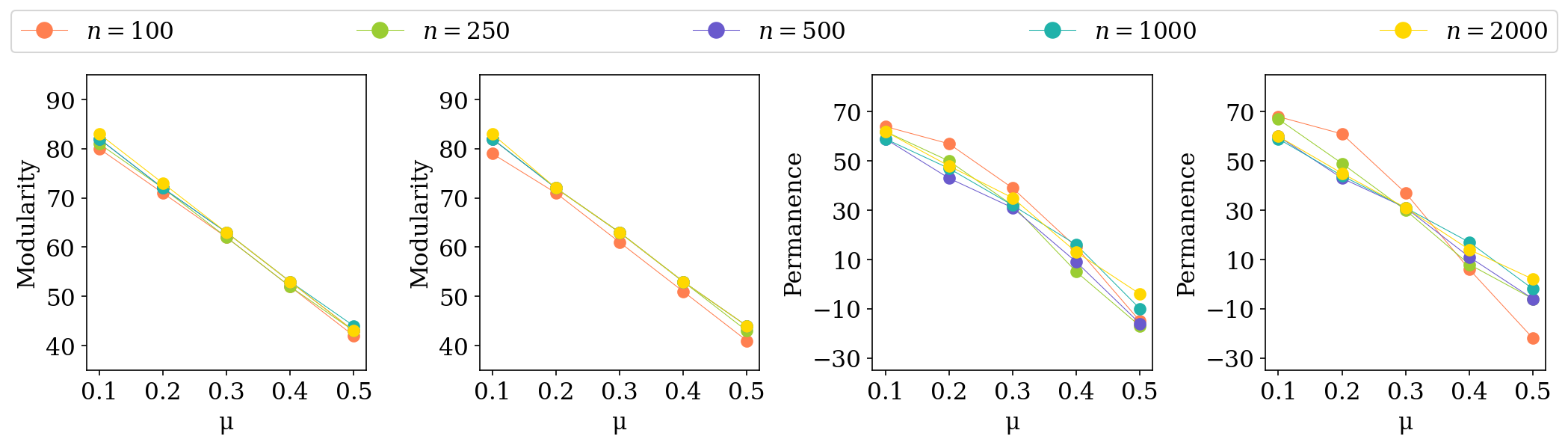} \\
~~~~~{(a)}\hspace{\foursep}{(b)}\hspace{\foursep}{(c)}\hspace{\foursep}{(d)} \\
\caption{The modularity with respect to the ground truth of the synthetic networks produced with the LFR library linearly depends on $\mu$ while is independent from $n$. Similarly, the permanence also exhibits a linear dependency on $\mu$ while is substantially independent from $n$. (a) Modularity plot with $\alpha=1/10$ and $\beta=30$. (b) Modularity plot with $\alpha=1/10$ and $\beta=40$. (c) Permanence plot with $\alpha=1/10$ and $\beta=30$. (d) Permanence plot with $\alpha=1/10$ and $\beta=40$.}
\label{fi:synthetic-ground-truth-varying-n}
\end{figure*}

\begin{itemize}
    \item Number of vertices $n \in \{100, 250, 500,$ $1000, 2000\}$
    \item Maximum vertex degree $d_{max} = \alpha \cdot n$ with $\alpha \in \{\frac{1}{20}, \frac{1}{15}, \frac{1}{10}, \frac{1}{5}, \frac{1}{3}\}$.
    \item Average vertex degree $\langle d \rangle = \beta \cdot d_{max} \cdot \frac{\log_{10}(n)}{n}$, with $\beta \in \{30, 35, 40\}$. Observe that, since  $d_{max} = \alpha \cdot n$, this corresponds to having $\langle d \rangle \sim \log(n)$. For example, when $\alpha=1/20$ and $\beta=30$, then $\langle d \rangle$ varies from $3$ ($n=100$) to $5$ ($n=2000)$. Instead, when $\alpha=1/3$ and $\beta=40$, then $\langle d \rangle$ varies from $26$ ($n=100$) to $44$ ($n=2000)$.
    \item Noise $\mu \in \{0.1, 0.2, 0.3, 0.4, 0.5\}$. 
\end{itemize}

\noindent This choice of the parameters yields networks of guaranteed modularity with respect to the ground truth's communities. In particular, as it can be seen from \Cref{fi:synthetic-ground-truth-varying-n}(a)-(b), the modularity of the produced instances for a given pair $\{\alpha, \beta\}$ only depends linearly on $\mu$ and is independent of the size~$n$ of the graph. The measure of the permanence with respect to the ground truth also seems to show a similar linear dependency on $\mu$ and a substantial independence from $n$ (see \Cref{fi:synthetic-ground-truth-varying-n}(c)-(d)).

\begin{figure*}[tb]
\centering
\includegraphics[width=0.80\linewidth]{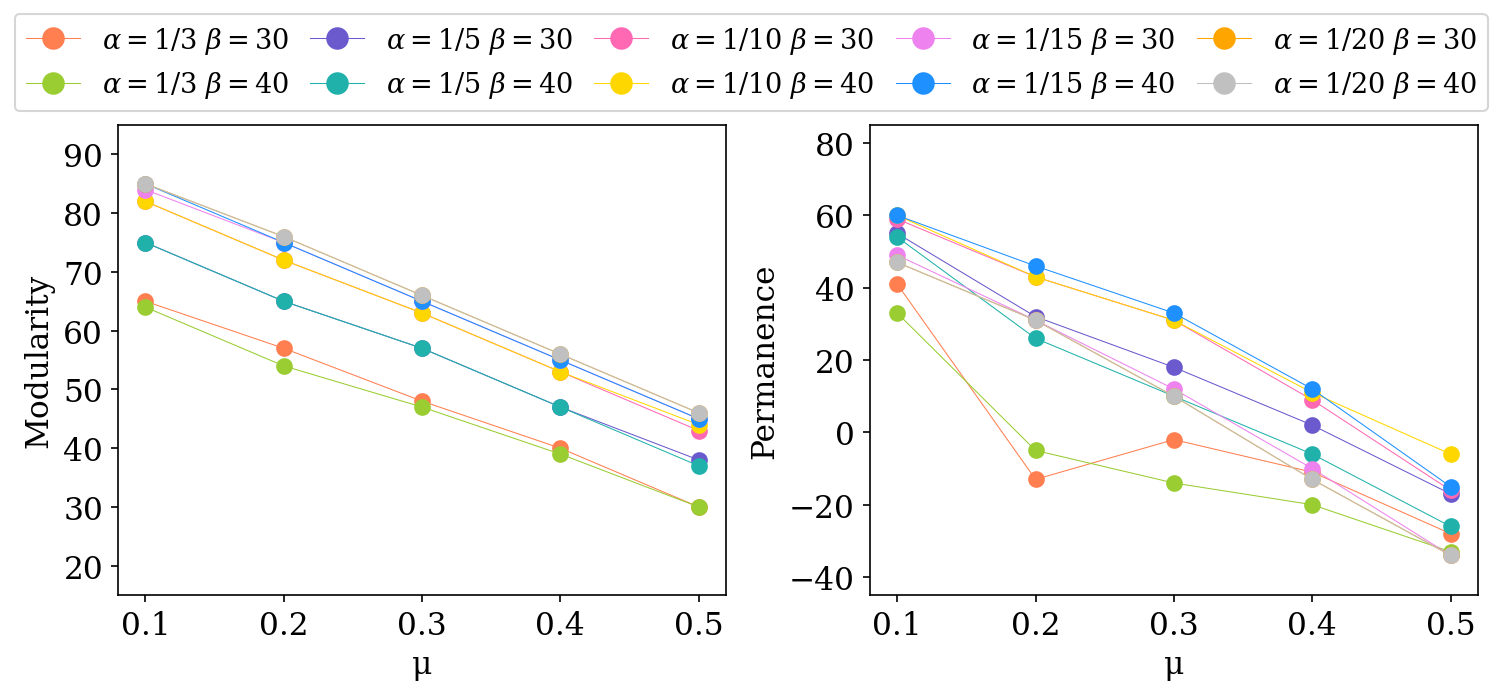} \\
~~~~~{(a)}\hspace{\twosep}{(b)} \\
\caption{The modularity and permanence with respect to the ground truth of the synthetic networks produced with the LFR library for different values of $\alpha$ and $\beta$. (a) Plot of modularity for $n=500$. The four curves for $\alpha \in \{1/15,1/20\}$ are all overlapped at the top of the plot. (b) Plot of permanence for $n=500$.} 
\label{fi:synthetic-ground-truth-varying-alpha-beta}
\end{figure*}

\cref{fi:synthetic-ground-truth-varying-alpha-beta}(a) shows that the modularity of the produced instances with respect to the ground truth is independent on the values of $\beta$ (i.e., from the average degree $\langle d \rangle$). Instead, the modularity decreases when $\alpha$ increases (i.e., when the scale-freeness of the produced networks increases). The measure of permanence (see \cref{fi:synthetic-ground-truth-varying-alpha-beta}(b)) shows a similar descending trend with respect to mixing parameter $\mu$, although more affected by noise. 

For each parameter setting, we generated five graph instances and report the average value of the considered metrics. The variance across instances was negligible and is therefore omitted from the figures for readability.

\subsection{Algorithms}
\label{ss:algorithms}

We evaluated our novel approach by comparing it with several state-of-the-art algorithms for network partitioning.
In particular, when the number of desired blocks is given, we compared in \cref{sec:experiments-fixed-k} with METIS~\cite{karypis1998fast}.
When the number of blocks of the partition is not known {\em a priori}, we compared in \cref{se:experiments} with Walktrap~\cite{https://doi.org/10.48550/arxiv.physics/0512106}, CNM~\cite{PhysRevE.70.066111}, Infomap~\cite{doi:10.1073/pnas.0706851105}, LP~\cite{Cordasco2010CommunityDV}, Pott~\cite{Reichardt_2006}, and Leiden~\cite{Leiden2019}. We were unable to compare with~\cite{DBLP:conf/nips/DevvritSD022} as the code is not yet publicly available.

As alternative network embedding approaches in the initial phase of {\em Clique-TF-IDF} we considered in the experimental analysis of \cref{sec:experiments-embedding-phase} 
algorithms Node2Vec~\cite{DBLP:conf/kdd/GroverL16} and DeepWalk~\cite{DBLP:conf/kdd/PerozziAS14}.
The source code of our implementation can be found at \url{https://github.com/mdelia17/clique-tf-idf}.

\section{Evaluation of {\em Clique-TF-IDF} for computing $k$-way partitions}\label{sec:experiments-fixed-k}

Assuming that the number $k$ of desired blocks of the partition is fixed, we evaluated the two variants of our approach called {\em $k$-Aggl-Clique-TF-IDF} and {\em $k$-Means-Clique-TF-IDF} (see \cref{se:methodology}) against algorithm METIS using multilevel multiway partitioning (METIS-M) and recursive bisection (METIS-R)~\cite{karypis1998fast}.
A comparison on real networks is not possible, since for most of them the ground-truth -- and hence the value of $k$ -- is not known in advance. We therefore focus on the synthetic benchmarks (see \cref{sss:synthetic}).

\subsection{Varying community interconnectedness by mixing parameter~$\mu$}
\begin{figure*}[tb]
\centering
\includegraphics[width=1\linewidth]{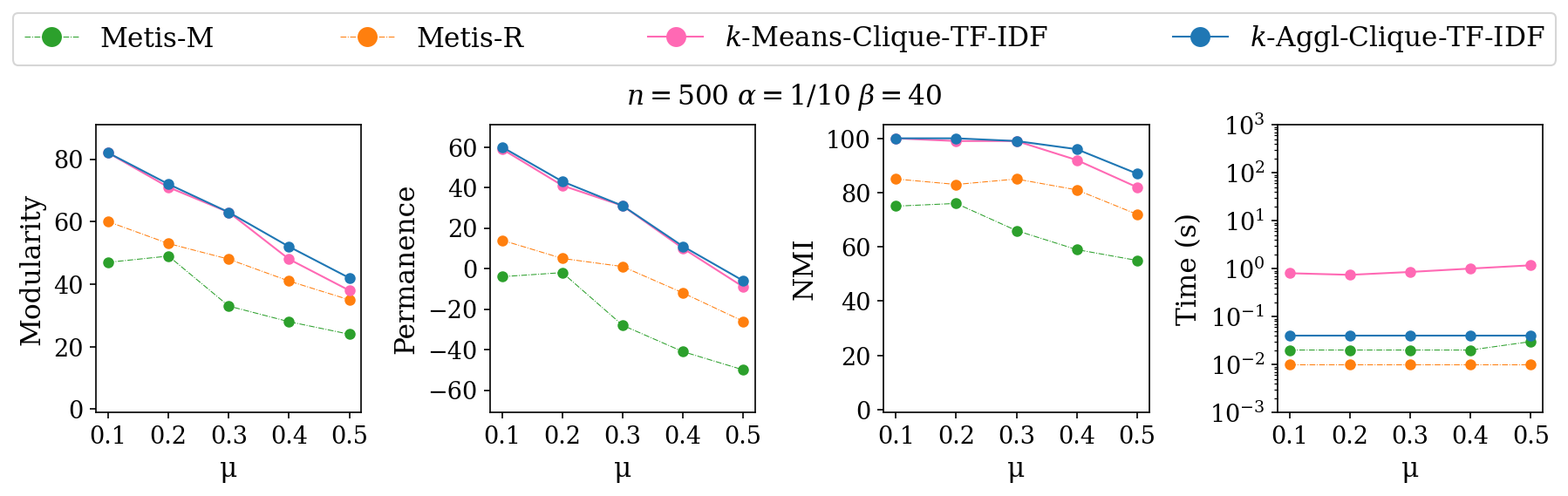} \\
~~~~~{(a)}\hspace{\foursep}{(b)}\hspace{\foursep}{(c)}\hspace{\foursep}{(d)} \\
 \includegraphics[width=1\linewidth]{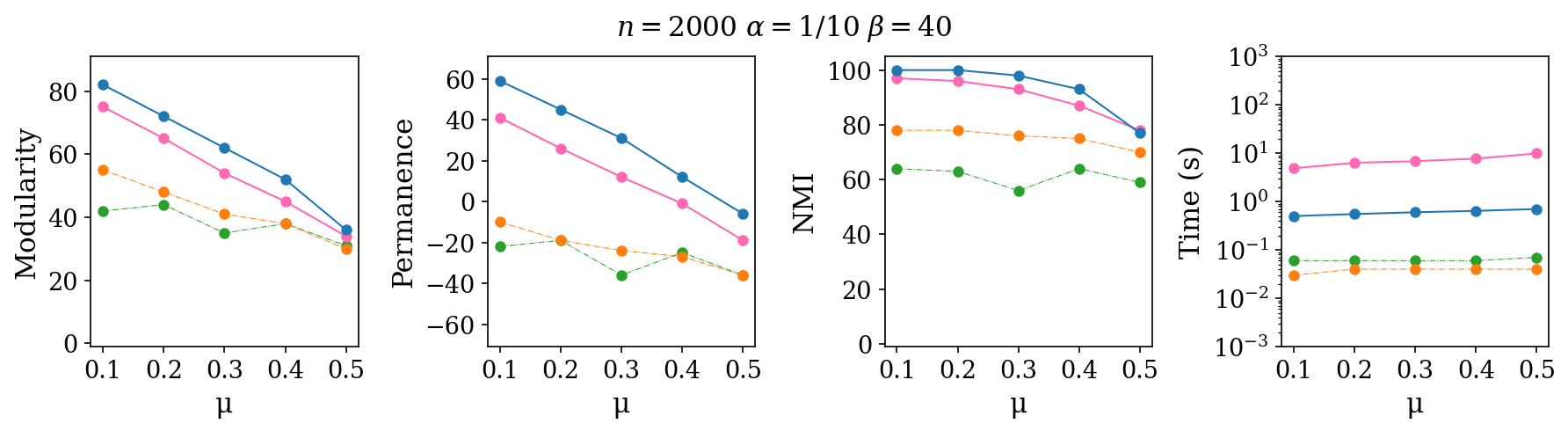} \\
~~~~~{(e)}\hspace{\foursep}{(f)}\hspace{\foursep}{(g)}\hspace{\foursep}{(h)} \\
\caption{Evaluation metrics on synthetic networks when $\alpha=1/10$ and $\beta=40$, varying $\mu$, for $n \in \{500,2000\}$ (rows).} 
\label{fi:k-clustering-synthetic-varying-mu}
\end{figure*}

We first analyze how the algorithms are affected by different values of the mixing parameter $\mu$. We fixed $\alpha=1/10$ and $\beta=40$. In \Cref{fi:k-clustering-synthetic-varying-mu}(a)-(c), when $n=500$, it can be seen that both our approaches lead in general to a better effectiveness in terms of modularity, permanence and NMI for all the values of $\mu$. Moreover, {\em $k$-Aggl-Clique-TF-IDF} has higher values of modularity and NMI with respect to {\em $k$-Means-Clique-TF-IDF} when $\mu > 0.3$.
In terms of efficiency, both METIS-M and METIS-R are always more efficient and {\em $k$-Means-Clique-TF-IDF} is one order of magnitude slower than {\em $k$-Aggl-Clique-TF-IDF} (see \cref{fi:k-clustering-synthetic-varying-mu}(d)). Furthermore, the running time of {\em $k$-Means-Clique-TF-IDF} slightly increase with $\mu$, while all the other algorithms do not exhibit a running time dependency on the noise $\mu$.

When $n=2000$, {\em $k$-Aggl-Clique-TF-IDF} remains the algorithm that produces higher quality communities, while {\em $k$-Means-Clique-TF-IDF} generally exhibits lower performances, even if higher than the other two algorithms (see \Cref{fi:k-clustering-synthetic-varying-mu}(e)-(g)). Furthermore, in terms of efficiency, the gap between the running time of {\em $k$-Aggl-Clique-TF-IDF} and both METIS algorithms raises, as depicted in \cref{fi:k-clustering-synthetic-varying-mu}(h).

\subsection{Varying the number of vertices}

\begin{figure*}[tb]
\centering
\includegraphics[width=1\linewidth]{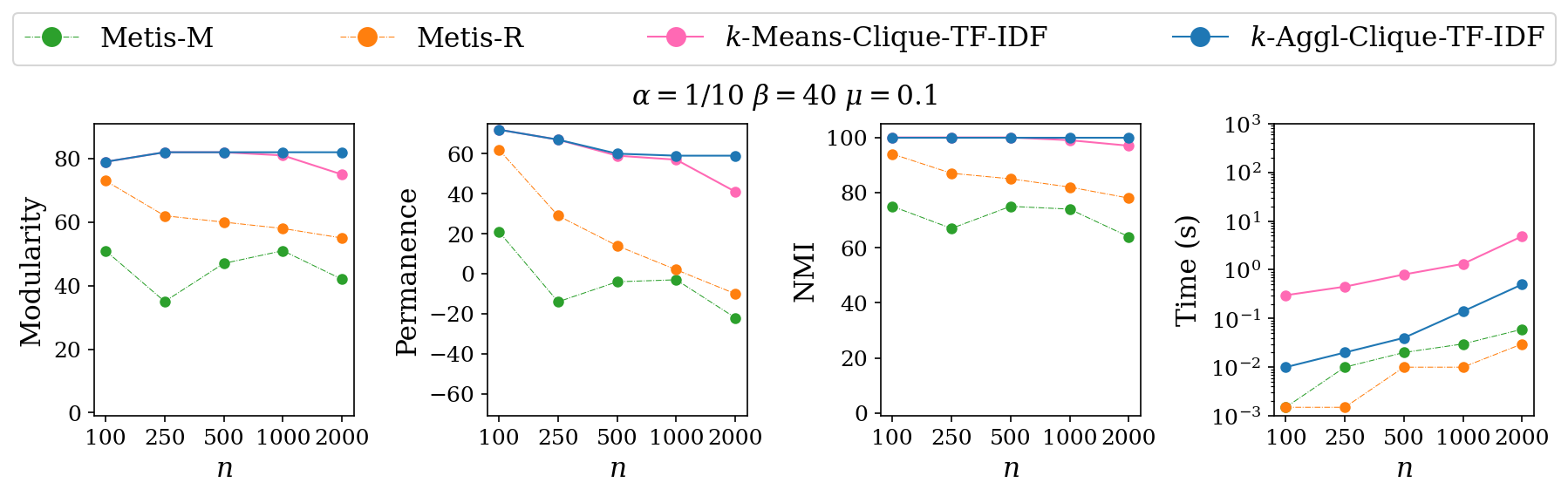} \\
~~~~~{(a)}\hspace{\foursep}{(b)}\hspace{\foursep}{(c)}\hspace{\foursep}{(d)} \\
 \includegraphics[width=1\linewidth]{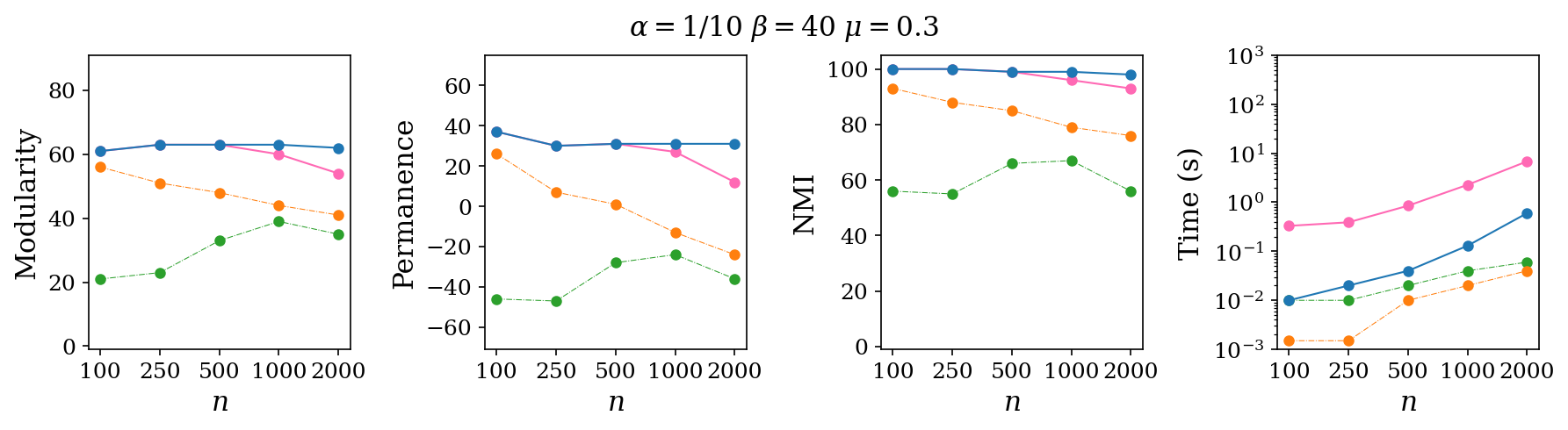} \\
~~~~~{(e)}\hspace{\foursep}{(f)}\hspace{\foursep}{(g)}\hspace{\foursep}{(h)} \\
\caption{Evaluation metrics on synthetic networks when $\alpha=1/10$ and $\beta=40$, varying $n$, for $\mu \in \{0.1,0.3\}$ (rows).} 
\label{fi:k-clustering-synthetic-varying-n}
\end{figure*}

We now evaluate the algorithms on all the evaluation metrics when the number of vertices $n$ in the network varies, while fixing $\alpha$ and $\beta$ to $1/10$ and $40$ respectively. \Cref{fi:k-clustering-synthetic-varying-n}(a)-(d) show results when $\mu=0.1$. As shown, both {\em $k$-Aggl-Clique-TF-IDF} and {\em $k$-Means-Clique-TF-IDF} exhibit better results in terms of quality, whereas both METIS-M and METIS-R generally struggle to discover qualitative communities as $n$ increases (see \Cref{fi:k-clustering-synthetic-varying-n}(a)-(c)). Moreover, {\em $k$-Aggl-Clique-TF-IDF} appears to be independent of $n$ in terms of both modularity and NMI.

\Cref{fi:k-clustering-synthetic-varying-n}(e)-(h) show the analysis when $\mu$ increases to $0.3$. It can be seen that {\em $k$-Aggl-Clique-TF-IDF} is still the best algorithm in terms of modularity and permanence (see \Cref{fi:k-clustering-synthetic-varying-n}(e)-(f)). Moreover, even with a higher value of the mixing parameter, our approach is able to discover communities that closely resemble those of the ground truth (see \cref{fi:k-clustering-synthetic-varying-n})(g).

Once again, the main limit of our approach is represented by an higher computation time and a stronger dependency from the number of vertices in the network, as it can be seen in \cref{fi:k-clustering-synthetic-varying-n}(d) and in \cref{fi:k-clustering-synthetic-varying-n}(h).

\subsection{Varying maximum vertex degree}
\begin{figure*}[tb]
\centering
\includegraphics[width=1\linewidth]{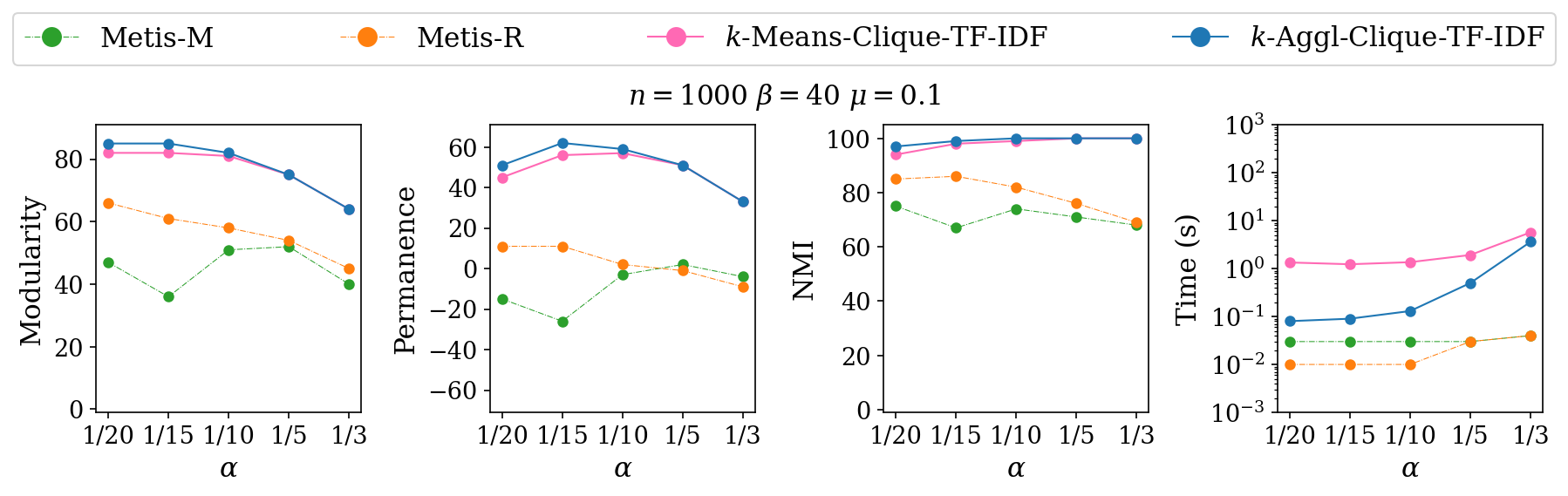} \\
~~~~~{(a)}\hspace{\foursep}{(b)}\hspace{\foursep}{(c)}\hspace{\foursep}{(d)} \\
 \includegraphics[width=1\linewidth]{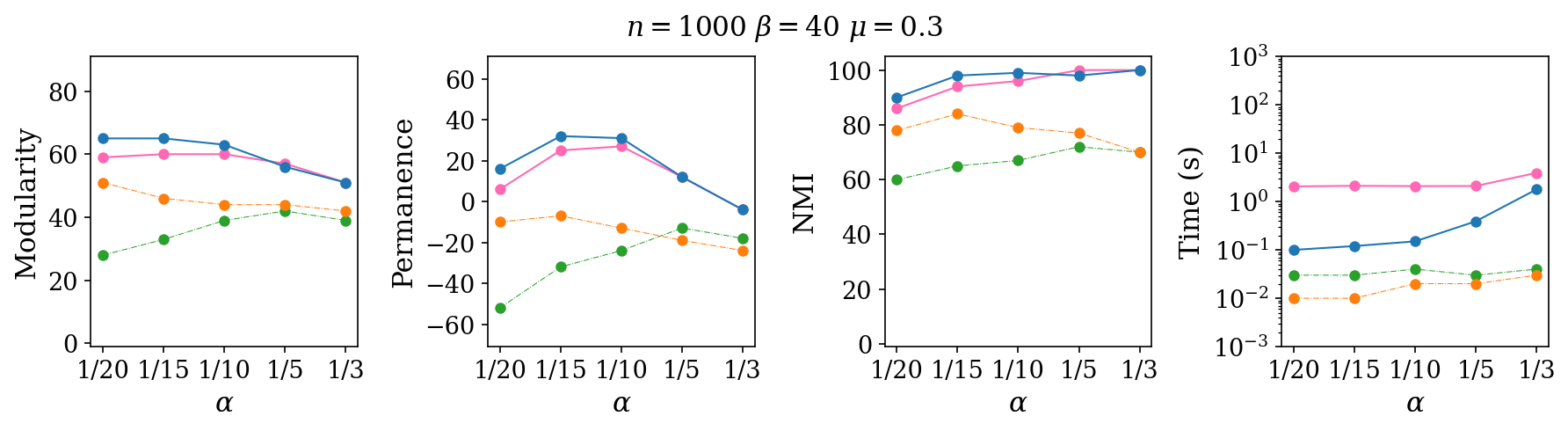} \\
~~~~~{(e)}\hspace{\foursep}{(f)}\hspace{\foursep}{(g)}\hspace{\foursep}{(h)} \\
\caption{Evaluation metrics on synthetic networks when $n=1000$ and $\beta=40$, varying $\alpha$, for $\mu \in \{0.1,0.3\}$ (rows).}  
\label{fi:k-clustering-synthetic-varying-alpha}
\end{figure*}

Finally, we report the evaluation with respect to different values of the maximum degree parameter $\alpha$. We fixed $n=1000$, $\beta=40$ and we considered $\mu=0.1$ (\Cref{fi:k-clustering-synthetic-varying-alpha}(a)-(d)) and $\mu=0.3$ (\Cref{fi:k-clustering-synthetic-varying-alpha}(e)-(h)).

In \Cref{fi:k-clustering-synthetic-varying-alpha}(a)-(b) and \Cref{fi:k-clustering-synthetic-varying-alpha}(e)-(f) it can be seen that both {\em $k$-Aggl-Clique-TF-IDF} and {\em $k$-Means-Clique-TF-IDF} are more effective with respect to METIS algorithms in terms of modularity and permanence, with {\em $k$-Aggl-Clique-TF-IDF} that appears to be slightly better with respect to {\em $k$-Means-Clique-TF-IDF} as $\alpha$ decreases. 
The partitioning that is returned by our approach is in general very close to the ground truth, while both METIS-M and METIS-R identify a partitioning that is far from it (see \cref{fi:k-clustering-synthetic-varying-alpha}(c) and \cref{fi:k-clustering-synthetic-varying-alpha}(g)).

\begin{figure}[h!]
\centering
\begin{tabular}{l c @{\hspace{0pt}} c}
& ~~~$\mu = 0.1$ & ~~~$\mu = 0.3$
\\

\begin{tabular}{l}
$\alpha=1/20$
\end{tabular}
& 
\begin{tabular}{l}
\includegraphics[width=0.30\linewidth]{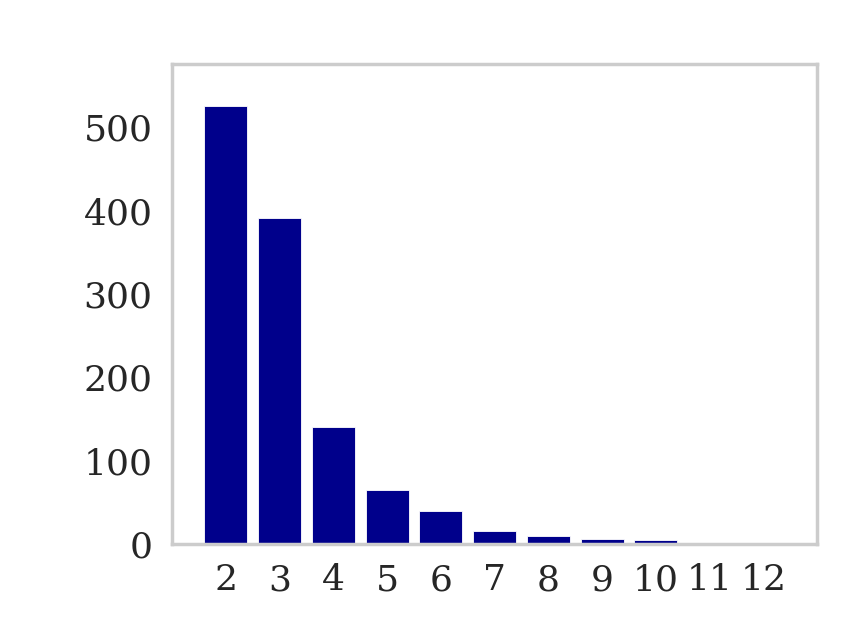}
\end{tabular}
&
\begin{tabular}{l}
\includegraphics[width=0.30\linewidth]{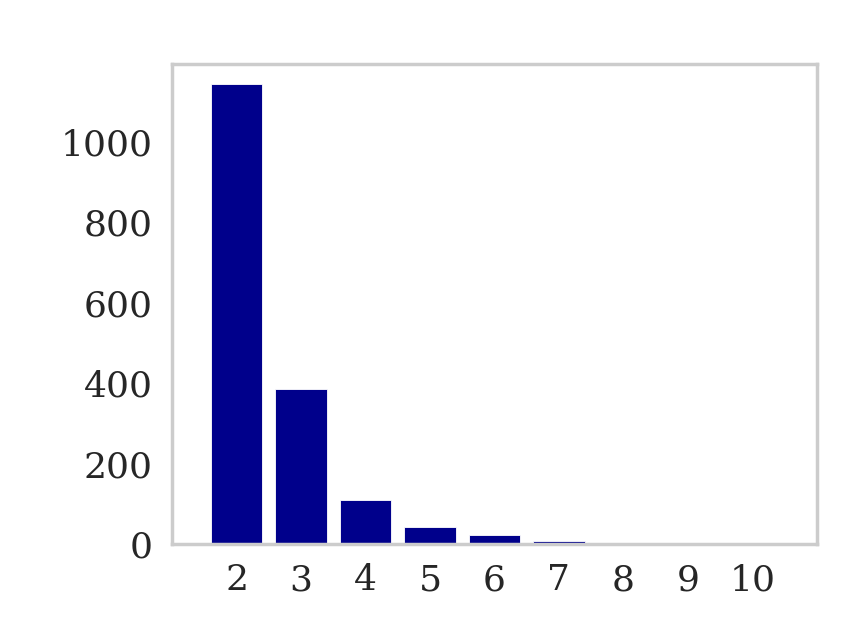}
\end{tabular}
\\

\begin{tabular}{l}
$\alpha=1/15$
\end{tabular}
& 
\begin{tabular}{l}
\includegraphics[width=0.30\linewidth]{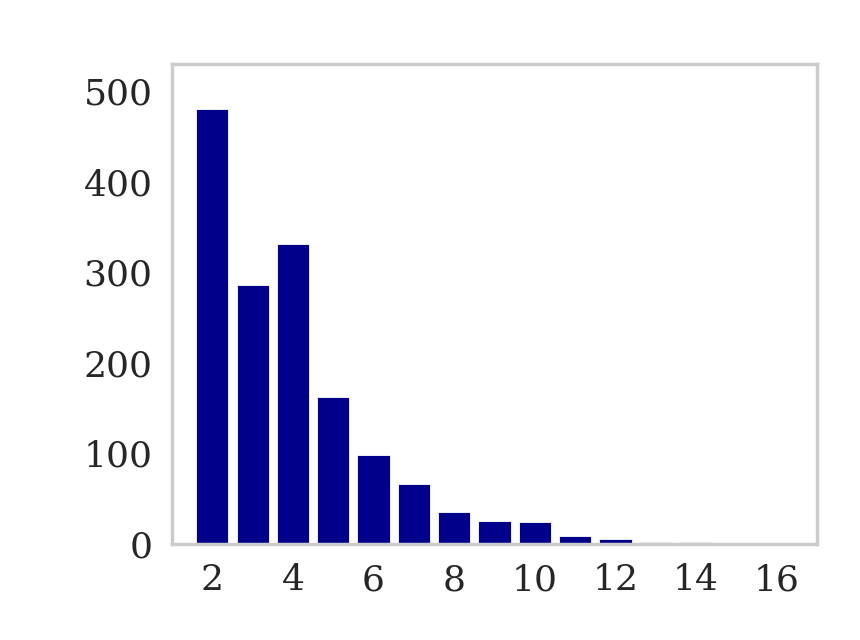}
\end{tabular}
&
\begin{tabular}{l}
\includegraphics[width=0.30\linewidth]{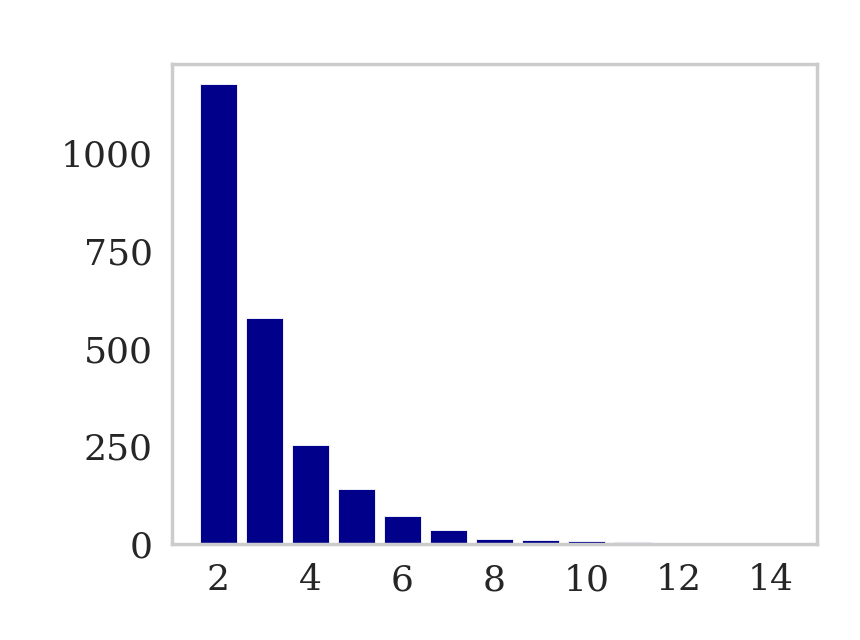}
\end{tabular}
\\

\begin{tabular}{l}
$\alpha=1/10$
\end{tabular}
& 
\begin{tabular}{l}
\includegraphics[width=0.30\linewidth]{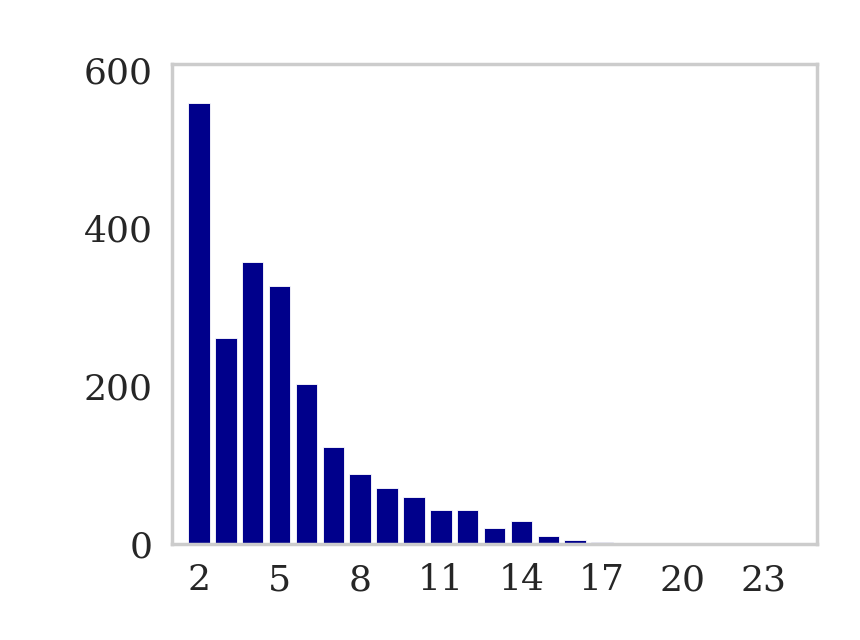}
\end{tabular}
&
\begin{tabular}{l}
\includegraphics[width=0.30\linewidth]{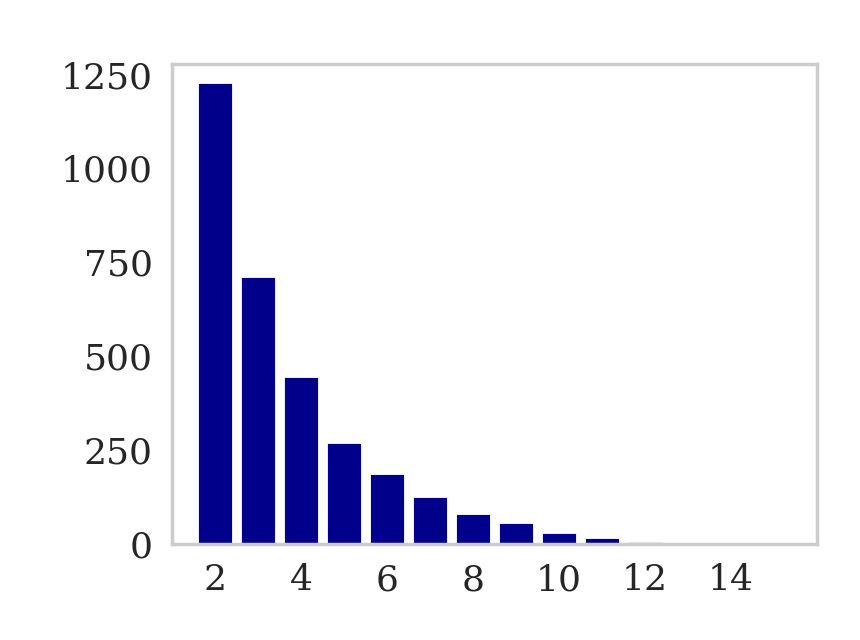}
\end{tabular}
\\

\begin{tabular}{l}
$\alpha=1/5$
\end{tabular}
& 
\begin{tabular}{l}
\includegraphics[width=0.30\linewidth]{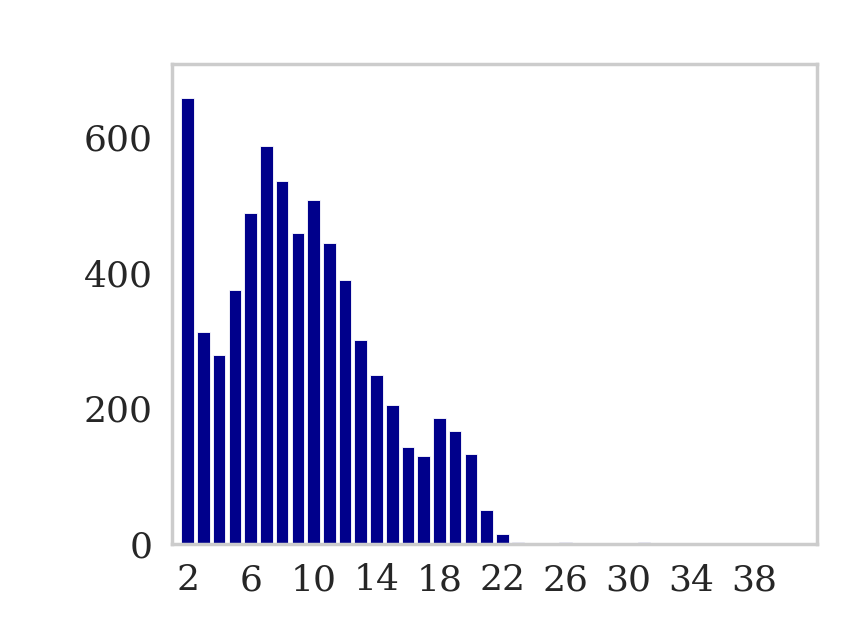}
\end{tabular}
&
\begin{tabular}{l}
\includegraphics[width=0.30\linewidth]{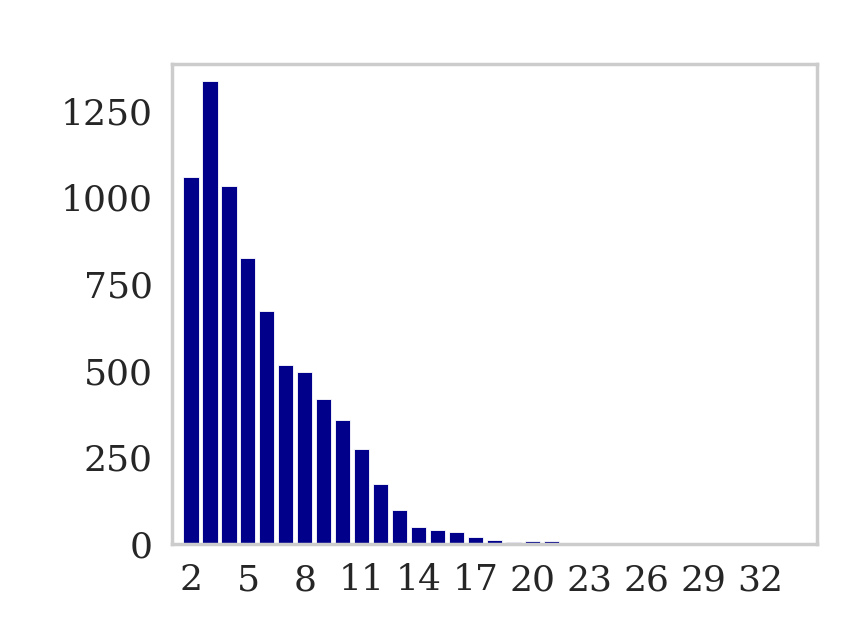}
\end{tabular}
\\

\begin{tabular}{l}
$\alpha=1/3$
\end{tabular}
& 
\begin{tabular}{l}
\includegraphics[width=0.30\linewidth]{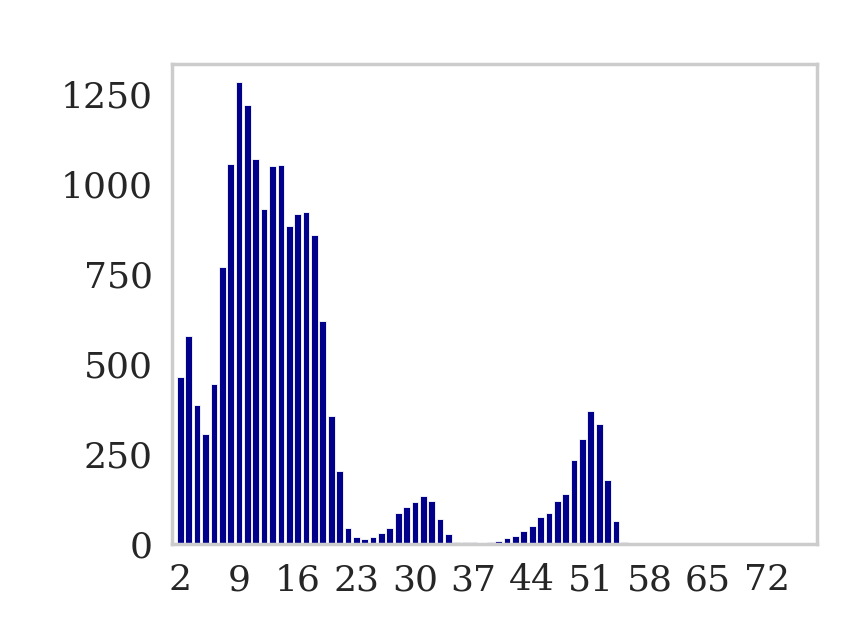}
\end{tabular}
&
\begin{tabular}{l}
\includegraphics[width=0.30\linewidth]{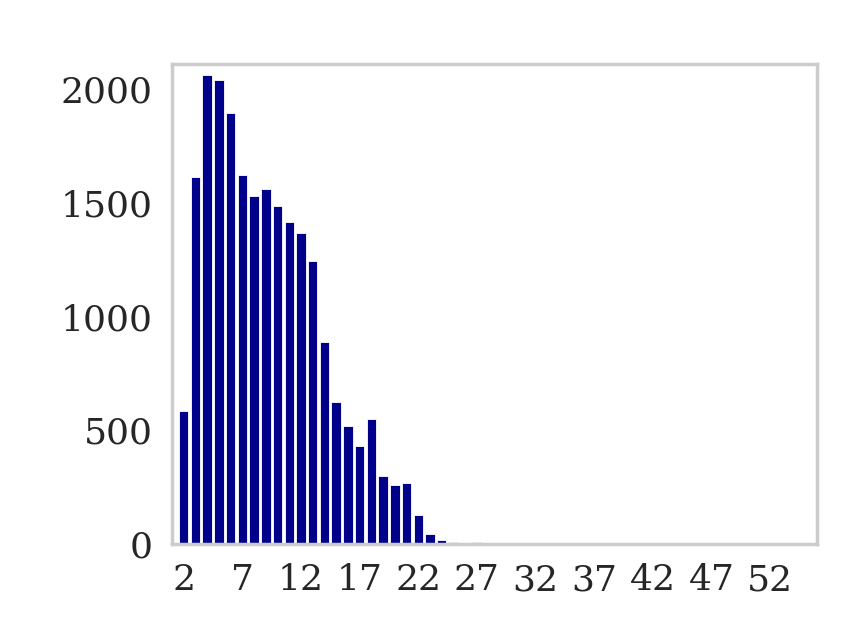}
\end{tabular}
\\

\end{tabular}
\caption{Distribution of the number of maximal cliques (y-axis) with respect to their size (x-axis) when $\beta=40$ and $n=1000$, for $\alpha \in \{\frac{1}{20}, \frac{1}{15}, \frac{1}{10}, \frac{1}{5}, \frac{1}{3}\}$ (rows) and for $\mu \in \{0.1,0.3\}$ (columns).} 
\label{fi:clique-distribution-synthetic-varying-alpha-and-mu}
\end{figure}

Moreover, \cref{fi:k-clustering-synthetic-varying-alpha}(d) and \cref{fi:k-clustering-synthetic-varying-alpha}(h) show that while {\em $k$-Means-Clique-TF-IDF} is the least efficient, {\em $k$-Aggl-Clique-TF-IDF} exhibits a stronger dependency of running time as $\alpha$ increases that is primarily attributed to the increased number of maximal cliques and their size. The distribution of the number of maximal cliques with respect to their size for different values of $\alpha$ is depicted in \cref{fi:clique-distribution-synthetic-varying-alpha-and-mu}. In particular, it can be seen that as $\alpha$ increases both the number of cliques and the size of the largest maximal clique increase.

\section{Evaluation of {\em Clique-TF-IDF} with unknown number of blocks}\label{se:experiments}

In this section we describe an experimental study to evaluate the effectiveness and efficiency of algorithm {\em Clique-TF-IDF}, when the number of blocks in the partition is not known {\em a priori}. 

\subsection{Experiments on real-world networks}\label{se:experiments-real}

\begin{figure*}[tb]
\centering
\includegraphics[width=1\linewidth]{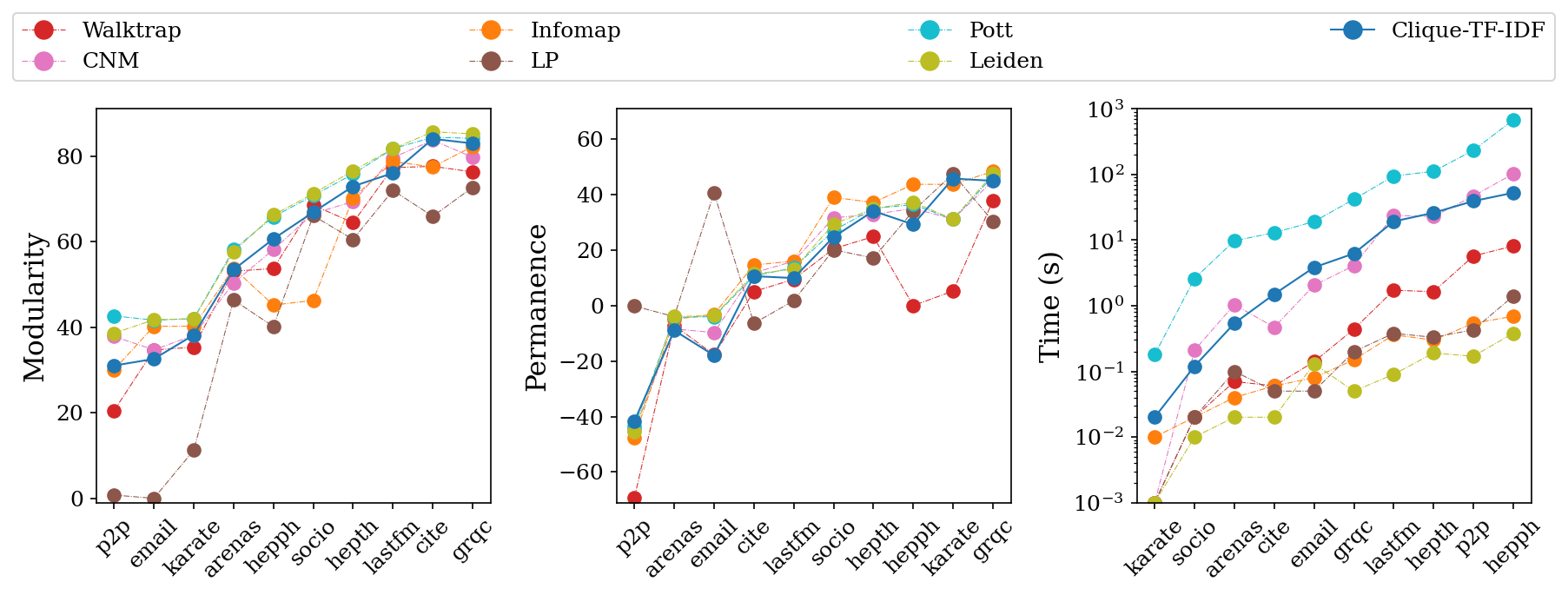} \\
~~~~~{(a)}\hspace{\threesep}{(b)}\hspace{\threesep}{(c)}
\caption{Effectiveness and efficiency measures achieved by each algorithm in terms of modularity (a), permanence (b), and running time in seconds (c). The networks on the x-axis are ordered based on the average modularity, on the average permanence, and on {\em Clique-TF-IDF} running time, respectively.} 
\label{fi:clustering-real-networks}
\end{figure*}

\begin{figure}[tb]
  \centering
  \includegraphics[width=0.42\linewidth]{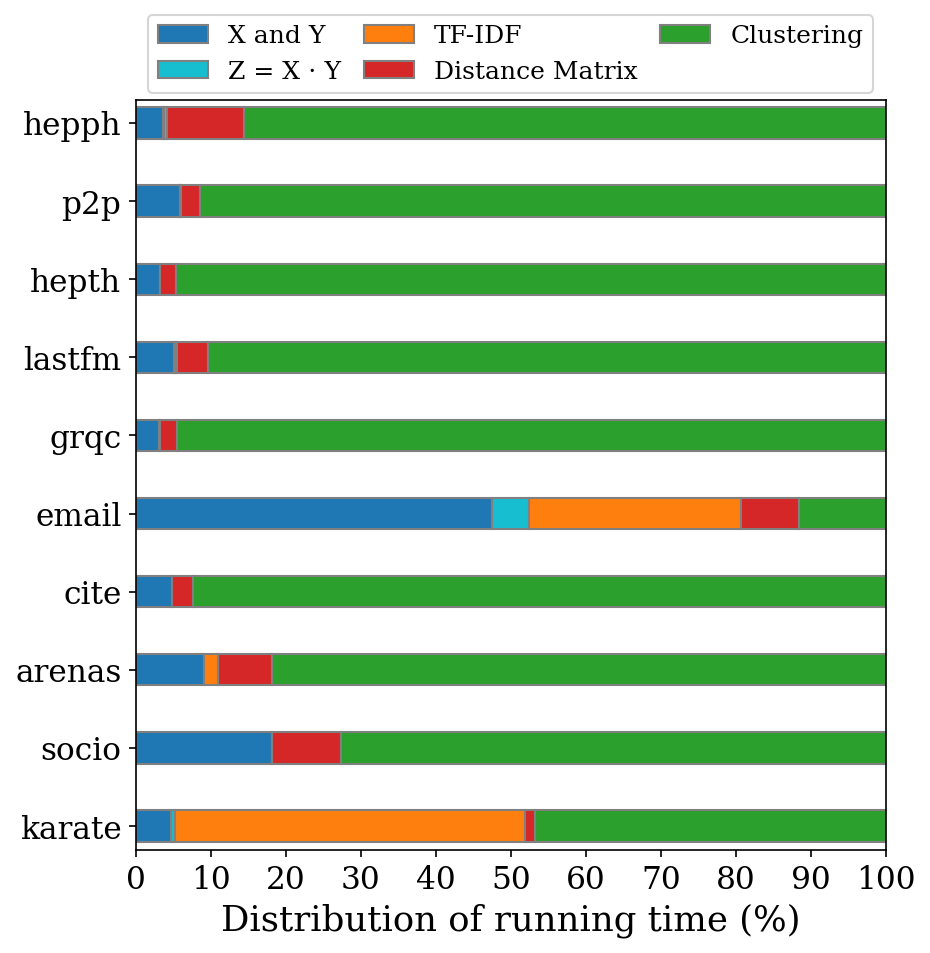}
  \caption{Distribution of the running time across each phase of {\em Clique-TF-IDF}.}
\label{fi:time-distribution}
\end{figure} 

We evaluate the efficiency and effectiveness of {\em Clique-TF-IDF} in comparison to the algorithms outlined in \cref{ss:algorithms}.
While {\em Clique-TF-IDF} being one of the algorithms that generates partitions with the highest modularity, Leiden consistently outperforms all others in terms of effectiveness (see \cref{fi:clustering-real-networks}(a)). Although Pott also achieves strong modularity results, it does so with significantly higher running times, at least an order of magnitude longer than its competitors.
Unlike its effectiveness on synthetic graphs (see \cref{sse:experiments-synthetic}), {\em Clique-TF-IDF} does not produce partitions with high permanence on real-world graphs compared to the other approaches (see \cref{fi:clustering-real-networks}(b)).

\begin{figure}
\centering
\begin{tabular}{c  c}
\includegraphics[width=0.38\linewidth]{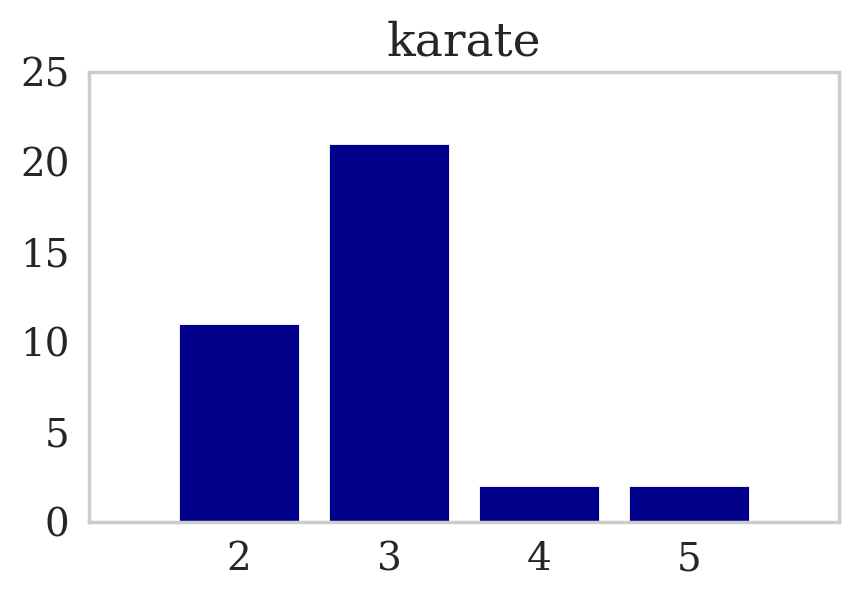} &
\includegraphics[width=0.38\linewidth]{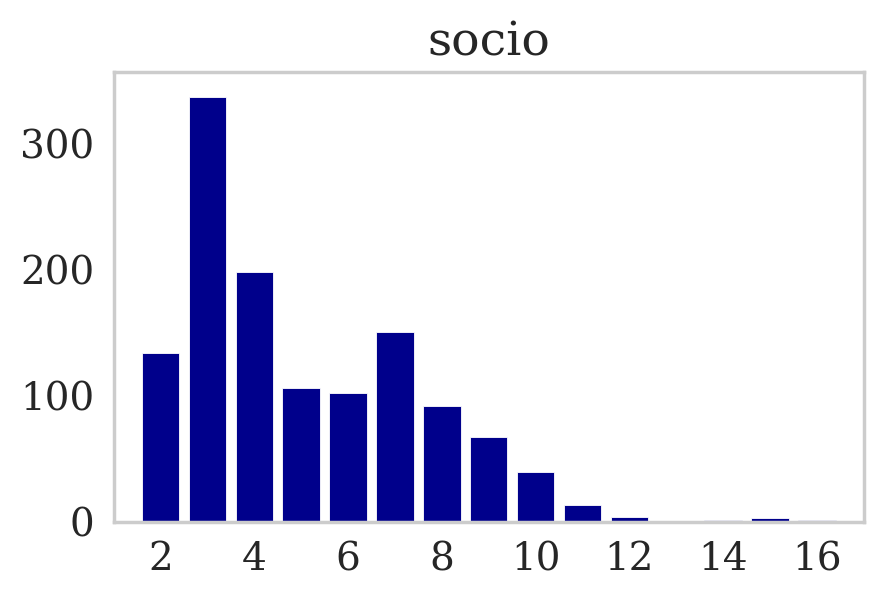} \\
\includegraphics[width=0.38\linewidth]{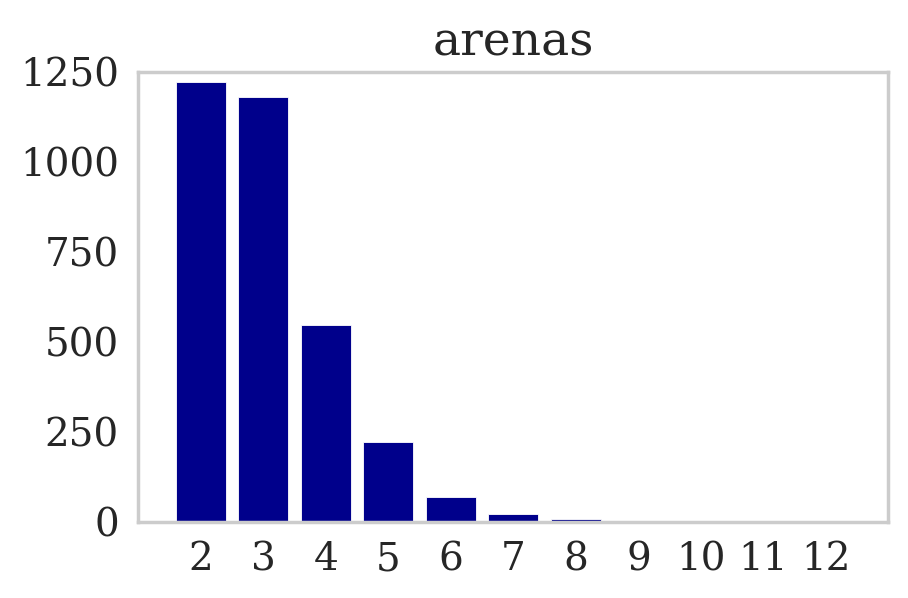} &
\includegraphics[width=0.38\linewidth]{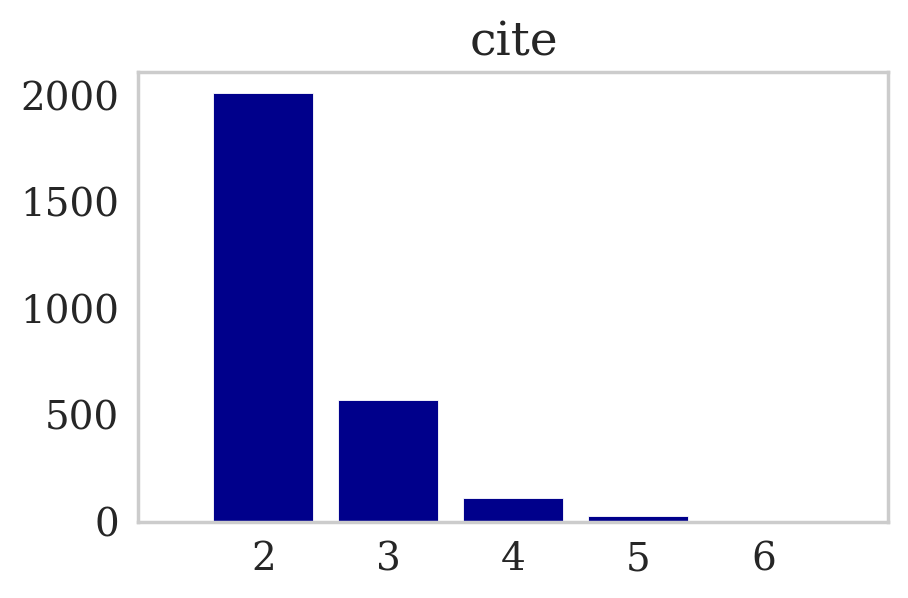} \\
\includegraphics[width=0.38\linewidth]{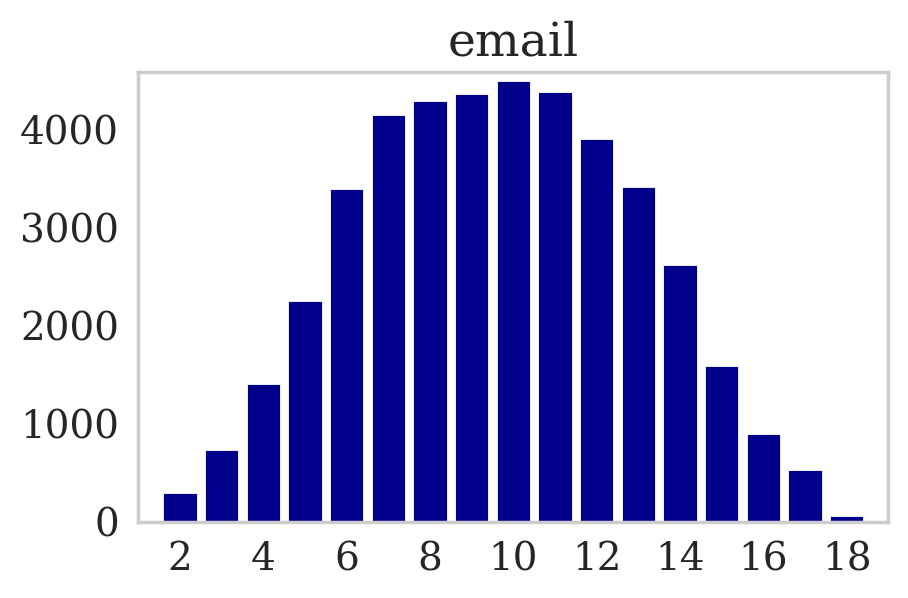} &
\includegraphics[width=0.38\linewidth]{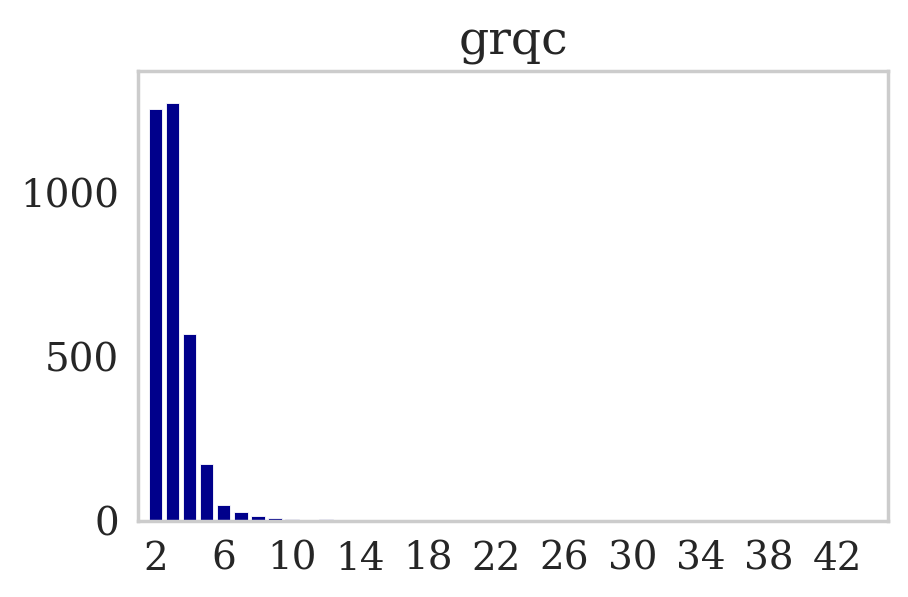} \\
\includegraphics[width=0.38\linewidth]{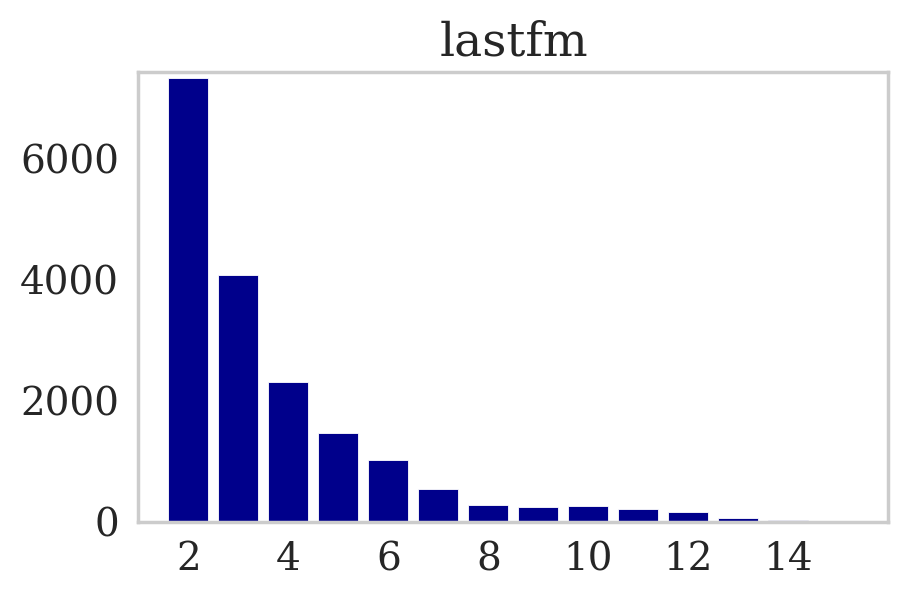} &
\includegraphics[width=0.38\linewidth]{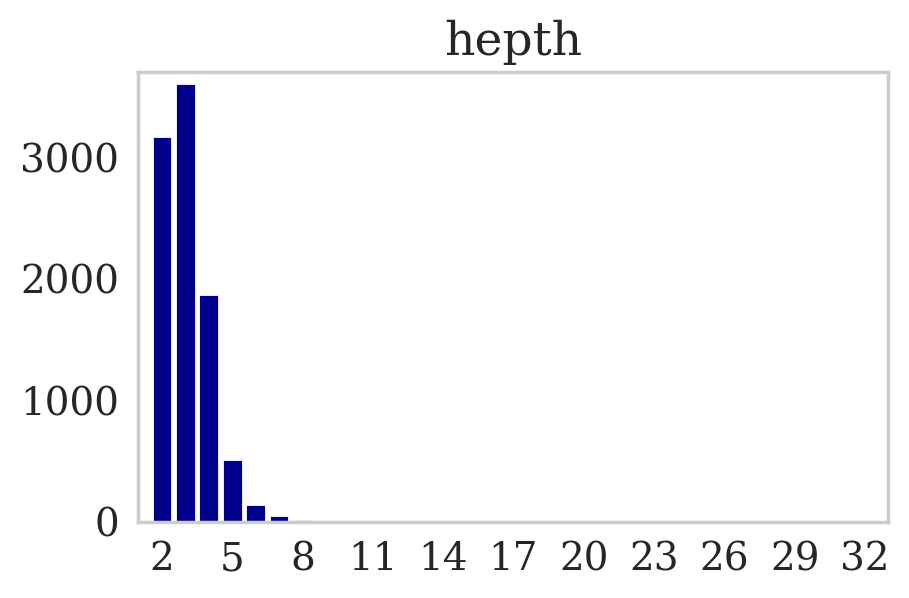} \\
\includegraphics[width=0.38\linewidth]{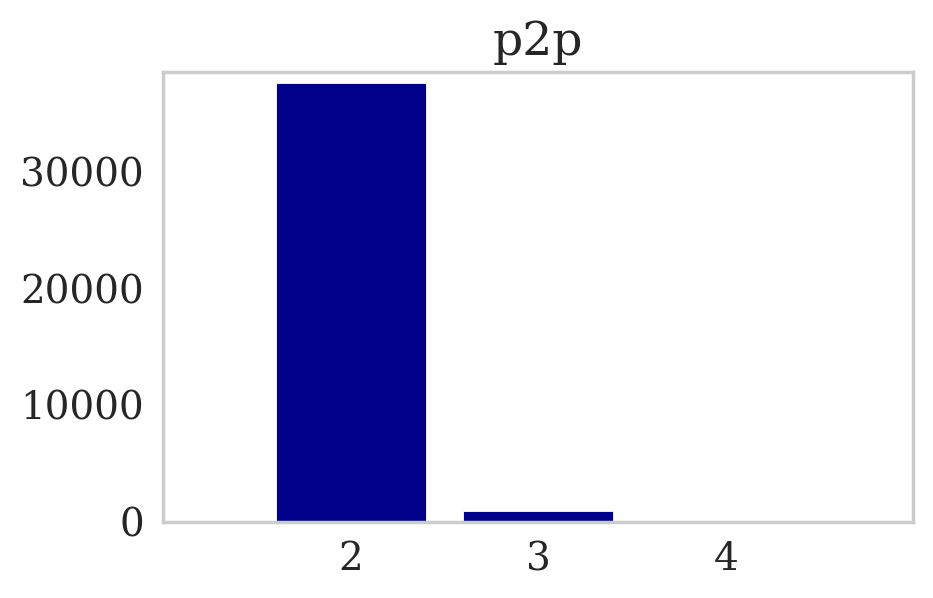} &
\includegraphics[width=0.38\linewidth]{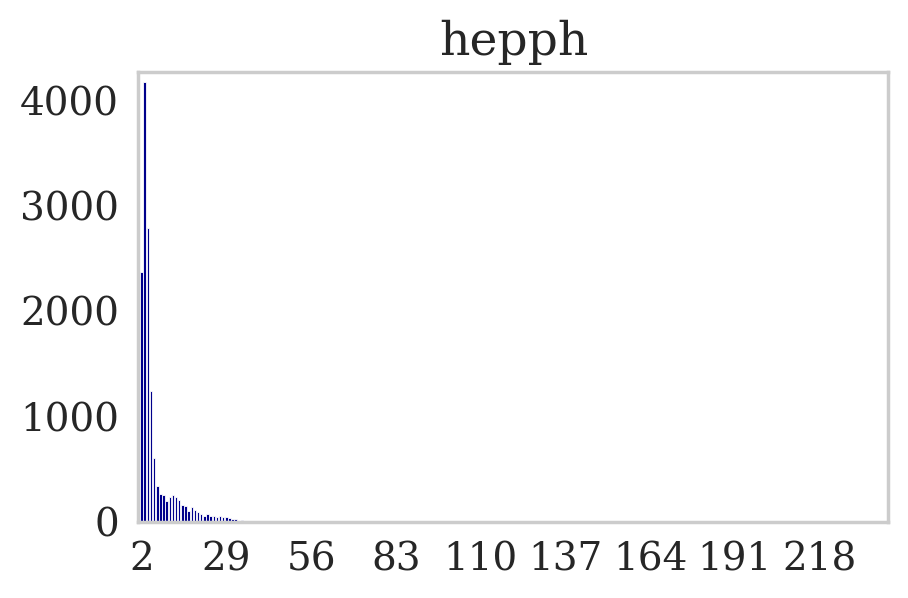} \\
\end{tabular}
\caption{Size distribution of maximal cliques for the real-world networks listed in \cref{ta:dataset}.}
\label{fi:clique-distribution}
\end{figure}

\cref{fi:clustering-real-networks}(c) illustrates that although {\em Clique-TF-IDF} algorithm has relatively long computation times, they remain lower than those of the Pott algorithm. The large number of maximal cliques in certain networks and the binary search for the optimal number $k$ of blocks are the primary causes of the high computation times. The overall runtime is dominated by the clustering phase, as revealed by a profiling analysis of {\em Clique-TF-IDF} (see \cref{fi:time-distribution}).
However, two networks stand out as exceptions to this trend, \texttt{sociopatterns} and, in particular, \texttt{email}. To explain this behavior, we examined the distribution of maximal cliques by size. \cref{fi:clique-distribution} reveals that these two networks have a large number of maximal cliques with a distribution that deviates from the typical scale-free pattern, resembling a binomial distribution in the case of \texttt{email}. As a result, a significant portion of time is spent on the computation of the $X$ and $Y$ matrices (blue rectangles in \cref{fi:time-distribution}).

\subsection{Experiments on synthetic networks}\label{sse:experiments-synthetic}
Here we discuss the evaluation of {\em Clique-TF-IDF} on synthetic graphs produced with LFR compared with other algorithms listed in \cref{ss:algorithms}. We analyze how modularity, permanence, NMI, and efficiency are affected by different values of the considered parameters.

\begin{figure*}[tb]
\centering
\includegraphics[width=1\linewidth]{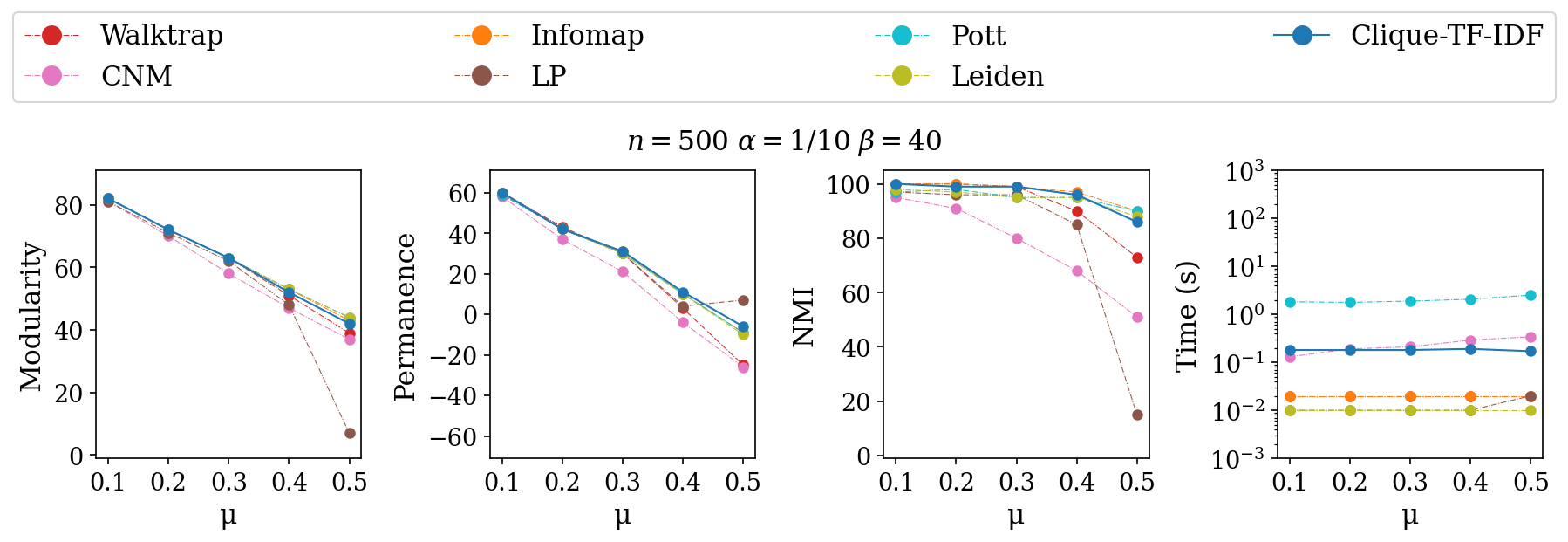} \\
~~~~~{(a)}\hspace{\foursep}{(b)}\hspace{\foursep}{(c)}\hspace{\foursep}{(d)} \\
 \includegraphics[width=1\linewidth]{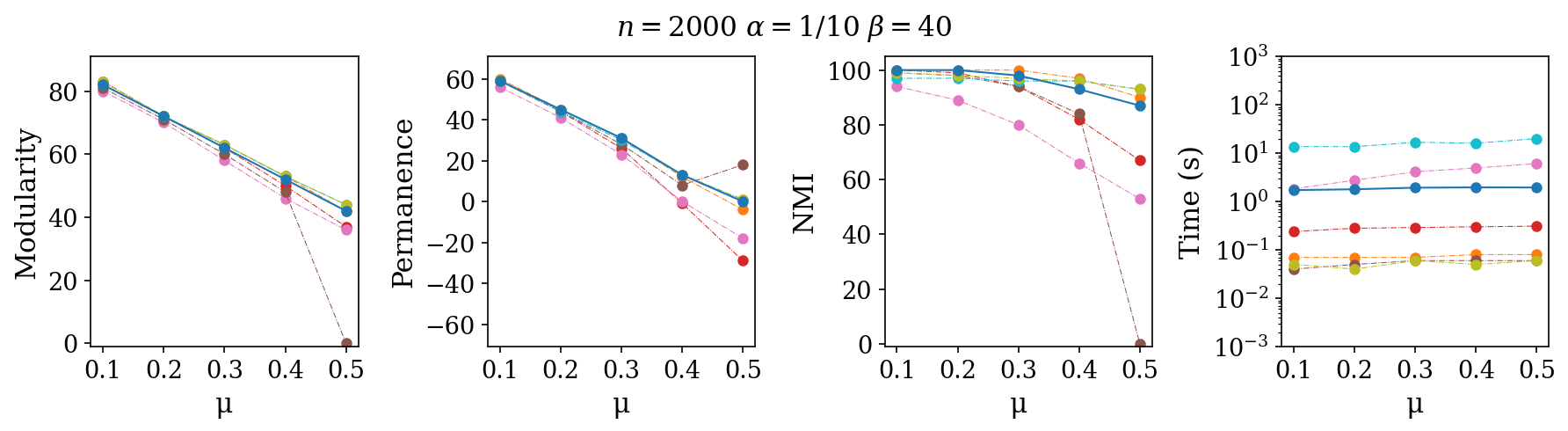} \\
~~~~~{(e)}\hspace{\foursep}{(f)}\hspace{\foursep}{(g)}\hspace{\foursep}{(h)} \\
\caption{Evaluation metrics on synthetic networks when $\alpha=1/10$ and $\beta=40$, varying $\mu$, for $n \in \{500,2000\}$ (rows).} 
\label{fi:clustering-synthetic-varying-mu}
\end{figure*}

\smallskip \noindent{\bf Varying community interconnectedness (mixing parameter~$\mu$).}
We fixed $\alpha=1/10$ and $\beta=40$, varying $\mu$ linearly in the interval $[0.1,0.5]$ for different values of $n$ to understand how the amount of noise in the network influences the ability of the algorithms to identify communities. As it can be seen in \Cref{fi:clustering-synthetic-varying-mu}(a)-(d) when $n=500$, {\em Clique-TF-IDF} is able to find blocks that are qualitative in terms of both modularity and permanence (see \Cref{fi:clustering-synthetic-varying-mu}(a)-(b)). More specifically, the modularity and permanence values of our approach are among the best for $\mu \leq 0.3$, while they are comparable for higher values of~$\mu$.

When comparing in terms of NMI with the ground truth offered by LFR (\cref{fi:clustering-synthetic-varying-mu}(c)), {\em Clique-TF-IDF} is able to maintain high NMI values also when the mixing parameter $\mu$ increases, while several algorithms like Walktrap, CNM, and LP decrease drastically their values of NMI when $\mu > 0.3$.
In terms of efficiency, our approach has among the largest computation times, that are however comparable with those of CNM and one order of magnitude smaller than those of algorithm Pott (see \cref{fi:clustering-synthetic-varying-mu}(d)).
Furthermore, from \cref{fi:clustering-synthetic-varying-mu}(d) it can be observed that {\em Clique-TF-IDF} exhibits an independence of its running times with respect to the mixing parameter, while the running times of some other algorithms like CNM slightly increase with $\mu$.
\Cref{fi:clustering-synthetic-varying-mu}(e)-(h) show an analogous analysis with $n=2000$, confirming that the above described behavior of the algorithms is substantially independent of the value of $n$ when the other parameters are fixed. 

\begin{figure*}[htb]
\centering
\includegraphics[width=1\linewidth]{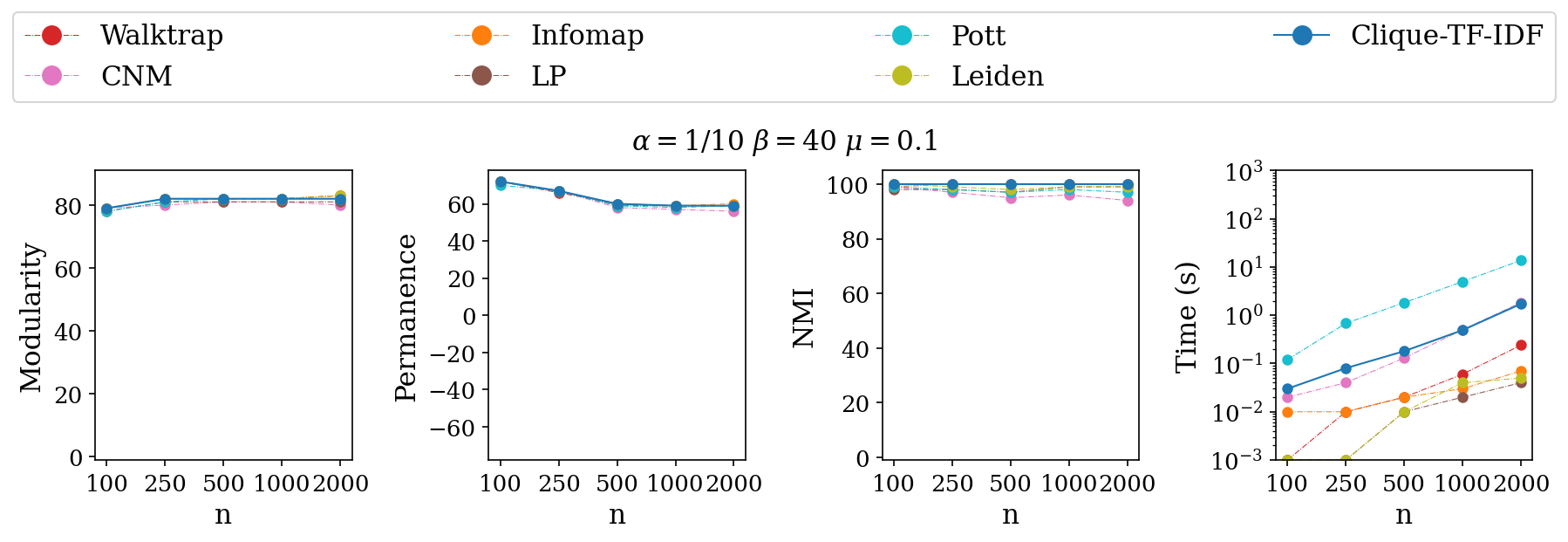} \\
~~~~~{(a)}\hspace{\foursep}{(b)}\hspace{\foursep}{(c)}\hspace{\foursep}{(d)} \\
 \includegraphics[width=1\linewidth]{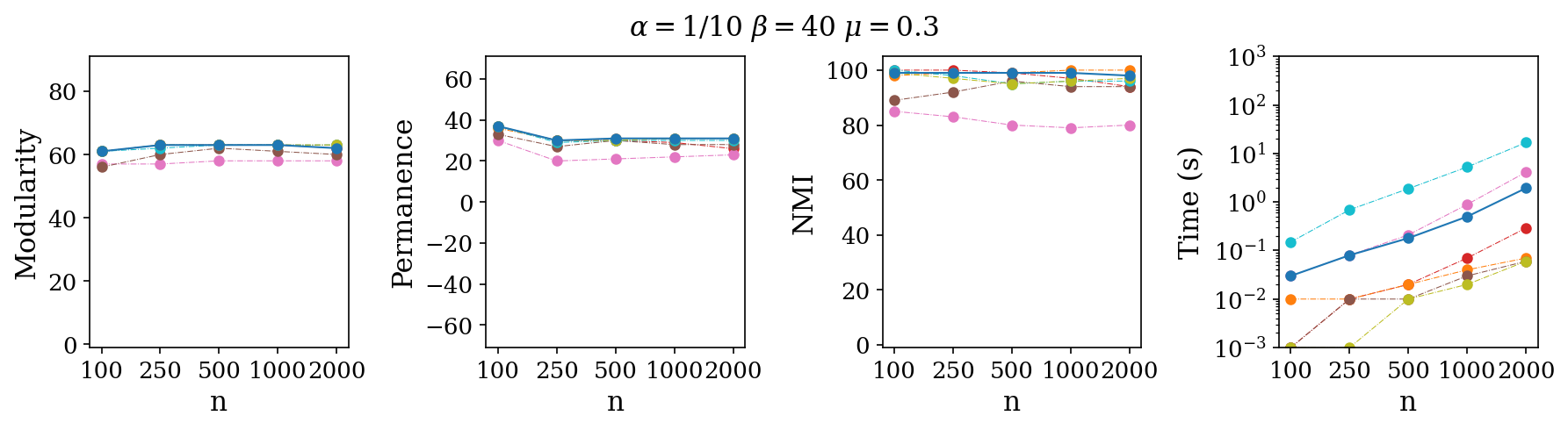} \\
~~~~~{(e)}\hspace{\foursep}{(f)}\hspace{\foursep}{(g)}\hspace{\foursep}{(h)} \\
\caption{Evaluation metrics on synthetic networks when $\alpha=1/10$ and $\beta=40$, varying $n$, for $\mu \in \{0.1,0.3\}$ (rows).} 
\label{fi:clustering-synthetic-varying-n}
\end{figure*}

\smallskip \noindent{\bf Varying the number of vertices.}
In the following, we analyze more in detail how the value of $n$ affects the four evaluation metrics. We fix $\alpha$ and $\beta$ respectively to $1/10$ and $40$, varying $n \in \{100,250,500,1000,2000\}$. \cref{fi:clustering-synthetic-varying-n} shows the analysis with the mixing parameter $\mu \in \{0.1,0.3\}$. In particular, it can be seen that {\em Clique-TF-IDF} has values of modularity, permanence, and NMI that are substantially independent of $n$, with curves that are always higher or comparable with the best algorithms in the state-of-the-art (see \Cref{fi:clustering-synthetic-varying-n}(a)-(c) and \Cref{fi:clustering-synthetic-varying-n}(e)-(g)). In terms of efficiency, our approach has the same dependency on $n$ of algorithm CNM (\cref{fi:clustering-synthetic-varying-n}(d) and \cref{fi:clustering-synthetic-varying-n}(h)). As in the case of real-world networks, algorithm Pott is the most inefficient.

\smallskip \noindent{\bf Varying maximum vertex degree.}
We fixed $n=1000$ and $\beta=40$ and evaluated how the performances of the algorithms vary when $\alpha$, i.e., the scale-freeness of the network, increases.

\begin{figure*}[tb]
\centering
\includegraphics[width=1\linewidth]{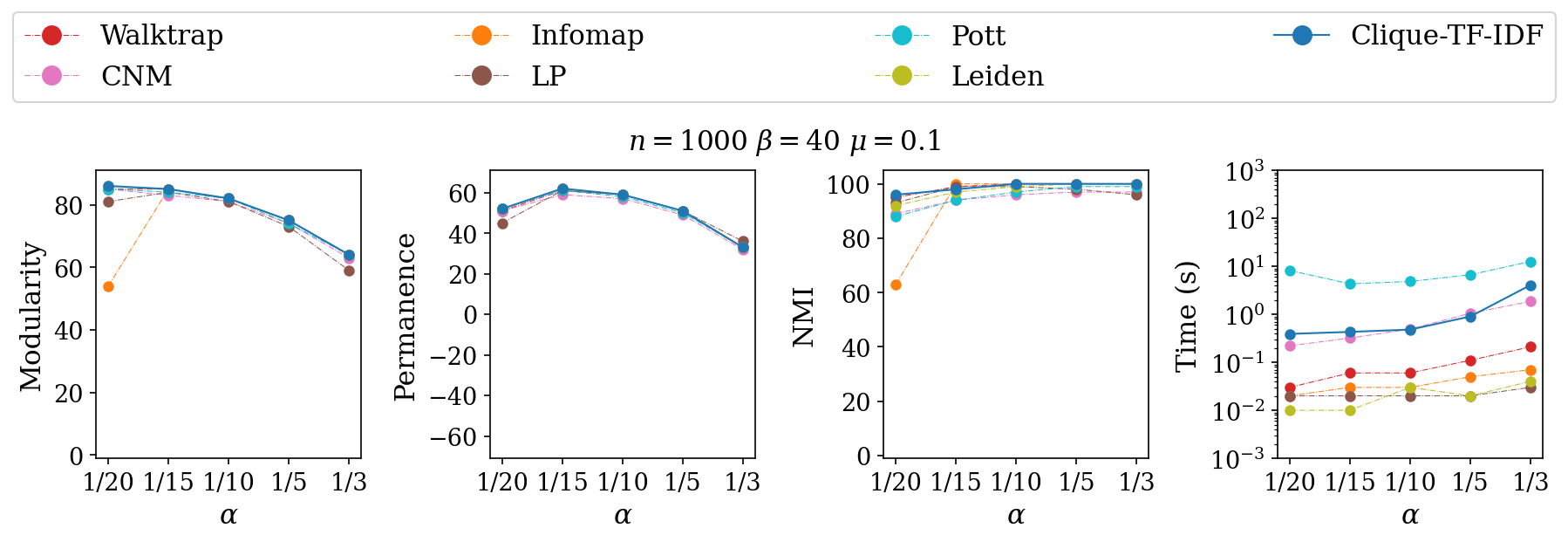} \\
~~~~~{(a)}\hspace{\foursep}{(b)}\hspace{\foursep}{(c)}\hspace{\foursep}{(d)} \\
 \includegraphics[width=1\linewidth]{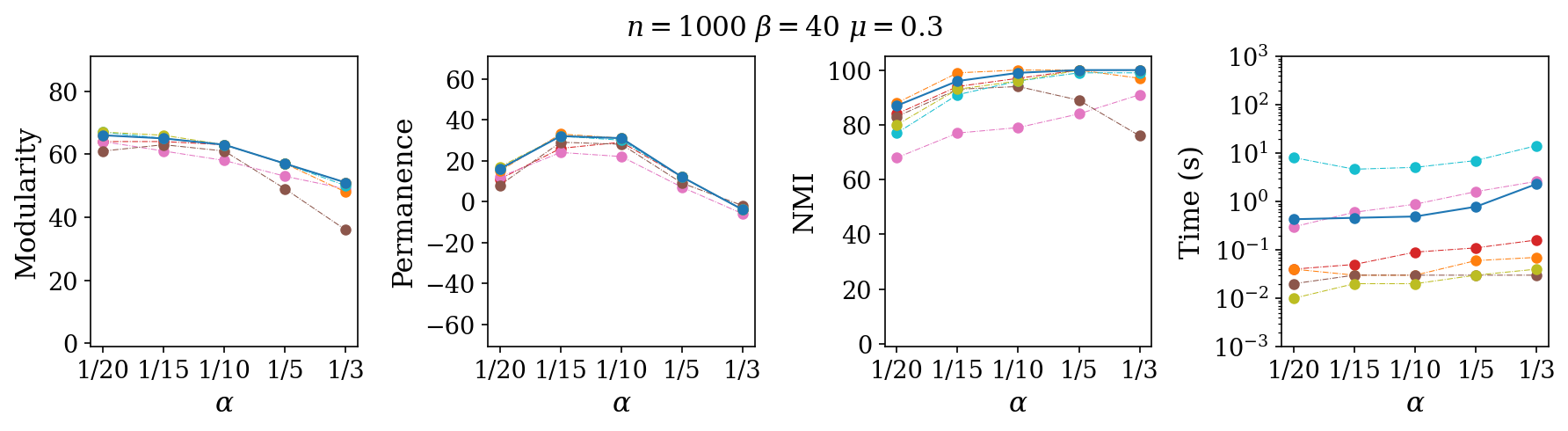} \\
~~~~~{(e)}\hspace{\foursep}{(f)}\hspace{\foursep}{(g)}\hspace{\foursep}{(h)} \\
\caption{Evaluation metrics on synthetic networks when $n=1000$ and $\beta=40$, varying $\alpha$, for $\mu \in \{0.1,0.3\}$ (rows).}  
\label{fi:clustering-synthetic-varying-alpha}
\end{figure*}

When $\mu=0.1$, in \Cref{fi:clustering-synthetic-varying-alpha}(a)-(c) it can be seen that {\em Clique-TF-IDF} can recover clusters that have in general the highest modularity and permanence. Additionally, our approach leads to a value of NMI that is higher and close to the maximum as $\alpha$ increases (i.e., the scale-freeness of the network increases), ensuring results among the best even for lower values of $\alpha$ (see \Cref{fi:clustering-synthetic-varying-alpha}(a)-(c)).

With regard to efficiency, \cref{fi:clustering-synthetic-varying-alpha}(d) shows that the running time of most algorithms increases as $\alpha$ grows. Notably, {\em Clique-TF-IDF} exhibits a stronger dependency, which can be attributed to the larger number of maximal cliques as it can be seen in \cref{fi:clique-distribution-synthetic-varying-alpha-and-mu}.

In \Cref{fi:clustering-synthetic-varying-alpha}(e)-(h)) it can be seen that our approach performs well in all quality metrics even when the noise increases, i.e., when $\mu=0.3$. In this case, when $\alpha > 1/10$, algorithm LP obtains lower values of modularity and NMI, while CNM has curves of modularity and NMI that are always below those of {\em Clique-TF-IDF}.
Although we don't include here the graphs, similar results are also obtained with higher values of the mixing parameter~$\mu$.

\section{Validating clique-based vertex embeddings}
\label{sec:experiments-embedding-phase}

The experiments discussed in this section aim at showing that the clique-based vertex embedding used in the pipeline of {\em Clique-TF-IDF} is a good starting point for computations involving communities in the graph. 
To achieve this, we applied our clustering phase to two popular graph embedding algorithms, Node2Vec~\cite{DBLP:conf/kdd/GroverL16} and DeepWalk~\cite{DBLP:conf/kdd/PerozziAS14}. Both algorithms produce positional vertex embeddings, where vertex similarity is based on adjacency in the input graph, making them suitable as a preliminary step for positional clustering, which is the focus of this paper.
We used the default values of DeepWalk for both the algorithms. This setting is denoted as D in \cref{fi:embedding-sota-eval,fi:embedding-synthetic-varying-mu,fi:embedding-synthetic-varying-n,fi:embedding-synthetic-varying-alpha}. Then, as these algorithms require many parameters in input, we considered different combinations of them. 
The number of random walks and epochs was fixed at 10, while we varied the embedding dimensionality $d$ within the set $d \in \{32, 64, 128\}$ and the random walk length $\ell$ within the set $\ell \in \{20, 40, 80\}$. All other parameters were left at their default values. In \cref{fi:embedding-sota-eval,fi:embedding-synthetic-varying-mu,fi:embedding-synthetic-varying-n,fi:embedding-synthetic-varying-alpha}, the settings labeled as S, M and L correspond to $d=32$ and $\ell=20$, $d=64$ and $\ell=40$, and $d=128$ and $\ell=80$, respectively.

\subsection{Real-world networks}

\begin{figure*}[tb]
\centering
\includegraphics[width=1\linewidth]{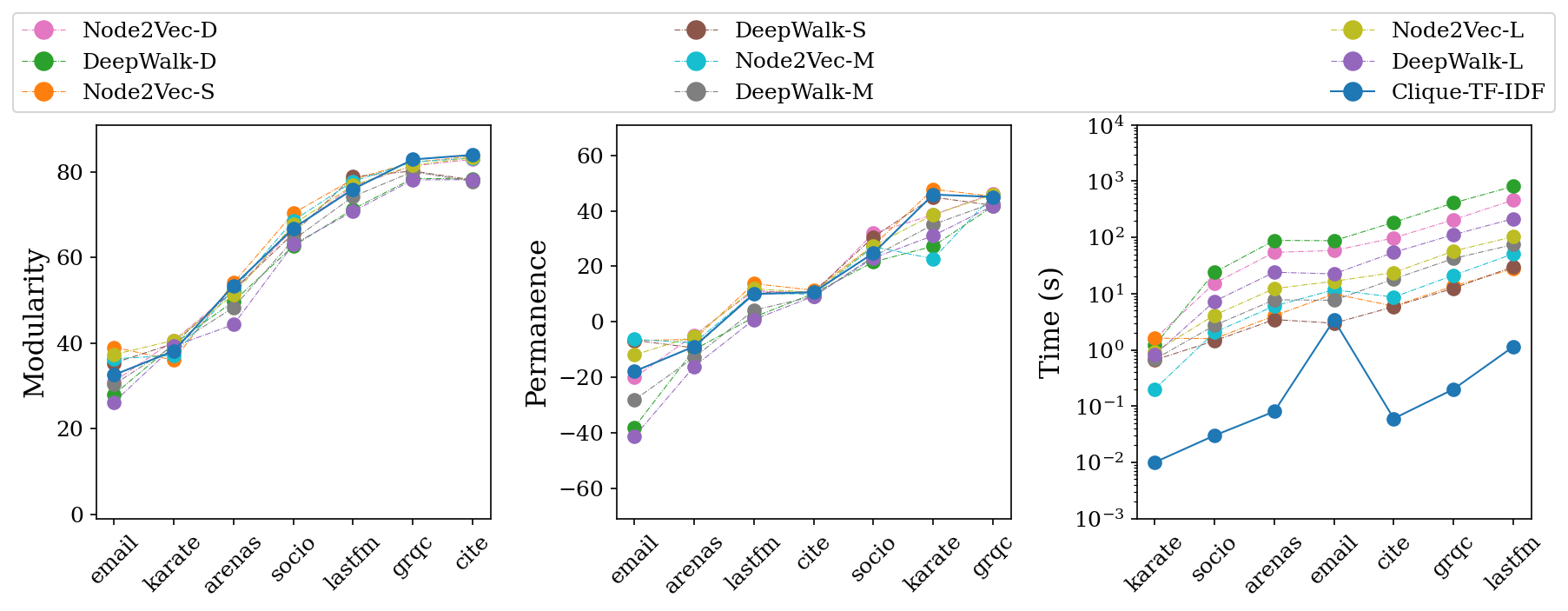} \\
~~~~~{(a)}\hspace{\threesep}{(b)}\hspace{\threesep}{(c)}
\caption{Effectiveness and efficiency measures achieved by DeepWalk,  Node2Vec and {\em Clique-TF-IDF} in terms of modularity (a), permanence (b), and running time in seconds (c). The networks on the x-axis are ordered based on the average modularity, on the average permanence, and on {\em Clique-TF-IDF} running time, respectively.}
\label{fi:embedding-sota-eval}
\end{figure*}

Since both Node2Vec and DeepWalk usually require substantial computation time, even on mid-sized networks, we considered only networks for which these algorithms can compute an embedding within a threshold time of $15$ minutes. The results of the embedding comparison, obtained using various configurations of Node2Vec and DeepWalk as discussed above, are shown in \cref{fi:embedding-sota-eval}. In terms of effectiveness, \cref{fi:embedding-sota-eval}(a) shows that the modularity of the partitions produced by {\em Clique-TF-IDF} is larger than that of DeepWalk in all of its settings, while it is comparable than that of Node2Vec (with the only exception of the graph \texttt{email}). Similar results can be observed in \cref{fi:embedding-sota-eval}(b) with respect to the permanence. Furthermore, variations in the parameters do not appear to significantly affect the quality of the partitions produced by the algorithms.

Moreover, as depicted in \cref{fi:embedding-sota-eval}(a)-(b), it is evident how there is no a clear winner across different settings of Node2Vec and DeepWalk, since the best choice of parameters may depend on the characteristics of the network. 
In terms of efficiency, \cref{fi:embedding-sota-eval}(c) shows that {\em Clique-TF-IDF} is significantly faster, by several orders of magnitude, than both DeepWalk and Node2Vec across all their settings (note that the y-axis of \cref{fi:embedding-sota-eval}(c) has a logarithmic scale). The increased computation time observed on \texttt{email} is due to the huge number of maximal clique in relation to the number of vertices.

\subsection{Synthetic networks}

Here we discuss the evaluation on synthetic graphs produced with LFR of {\em Clique-TF-IDF} compared with the four settings of Node2Vec and DeepWalk. We repeat the same analysis of \cref{sse:experiments-synthetic} on all the evaluation metrics.

\begin{figure*}[tb]
\centering
\includegraphics[width=1\linewidth]{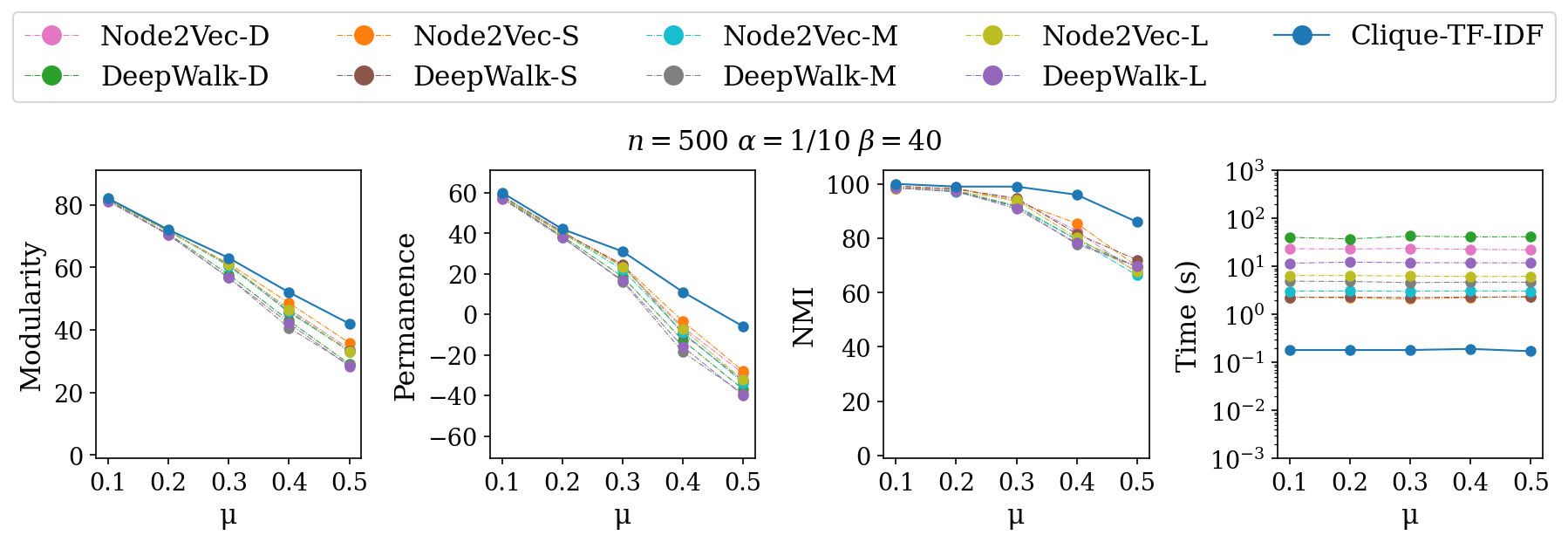} \\
~~~~~{(a)}\hspace{\foursep}{(b)}\hspace{\foursep}{(c)}\hspace{\foursep}{(d)} \\
 \includegraphics[width=1\linewidth]{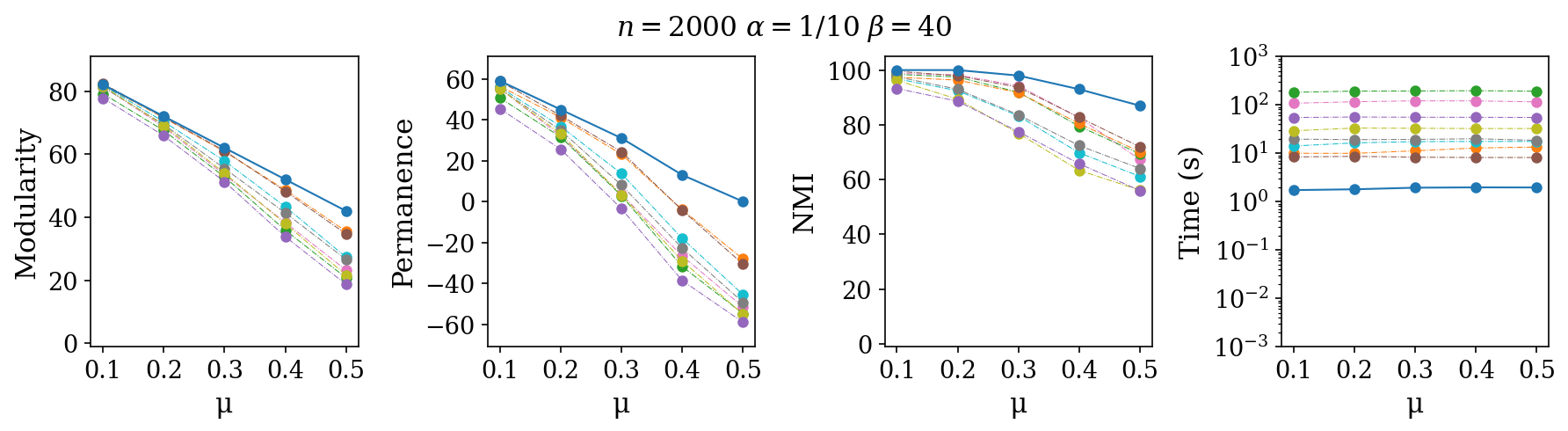} \\
~~~~~{(e)}\hspace{\foursep}{(f)}\hspace{\foursep}{(g)}\hspace{\foursep}{(h)} \\
\caption{Evaluation metrics on synthetic networks when $n=500, \alpha=1/10$ and $\beta=40$, varying $\mu$ linearly in the interval $[0.1,0.5]$.} 
\label{fi:embedding-synthetic-varying-mu}
\end{figure*}

\smallskip \noindent{\bf Varying community interconnectedness (mixing parameter~$\mu$).}
We first fixed $\alpha=1/10$ and $\beta=40$, varying $\mu$ linearly in the interval $[0.1,0.5]$ for different values of $n$, to understand how the amount of noise in the network influences the ability of the algorithms to discover strong communities. As it can be seen in \cref{fi:embedding-synthetic-varying-mu} with $n=500$, {\em Clique-TF-IDF} is able to find more qualitative blocks with respect to Node2Vec and DeepWalk in terms of both modularity and permanence (see \Cref{fi:embedding-synthetic-varying-mu}(a)-(b)). More specifically, the curves for modularity and permanence of Node2Vec and DeepWalk start to diverge from our curve when $\mu=0.3$, showing a clear sensitivity of the embedding quality produced by other algorithms as the noise value increases.

When comparing with the ground truth produced by LFR in terms of NMI (\cref{fi:embedding-synthetic-varying-mu}(c)), {\em Clique-TF-IDF} is able to discover most of it also when the mixing parameter $\mu$ grows, while other algorithms make more errors in classifying communities.
Our approach also outperforms the competitors in terms of efficiency, since we are able to produce an embedding in a time that is several order of magnitude lesser than those of Node2Vec and DeepWalk (see \cref{fi:embedding-synthetic-varying-mu}(d)). It can be observed as well that {\em Clique-TF-IDF} is substantially independent of the noise.

The same analysis is repeated when $n=2000$. In \Cref{fi:embedding-synthetic-varying-mu}(a)-(c), it can be seen that the curves for {\em Clique-TF-IDF} in terms of modularity, permanence and NMI are similar, so our approach is independent to the value of $n$ when other parameters are fixed. Instead, performance of other approaches tend to decrease when $n$ grows and this is a further motivation that parameters of these embedding algorithms have a strong dependency on the characteristics of the network.

\begin{figure*}[tb]
\centering
\includegraphics[width=1\linewidth]{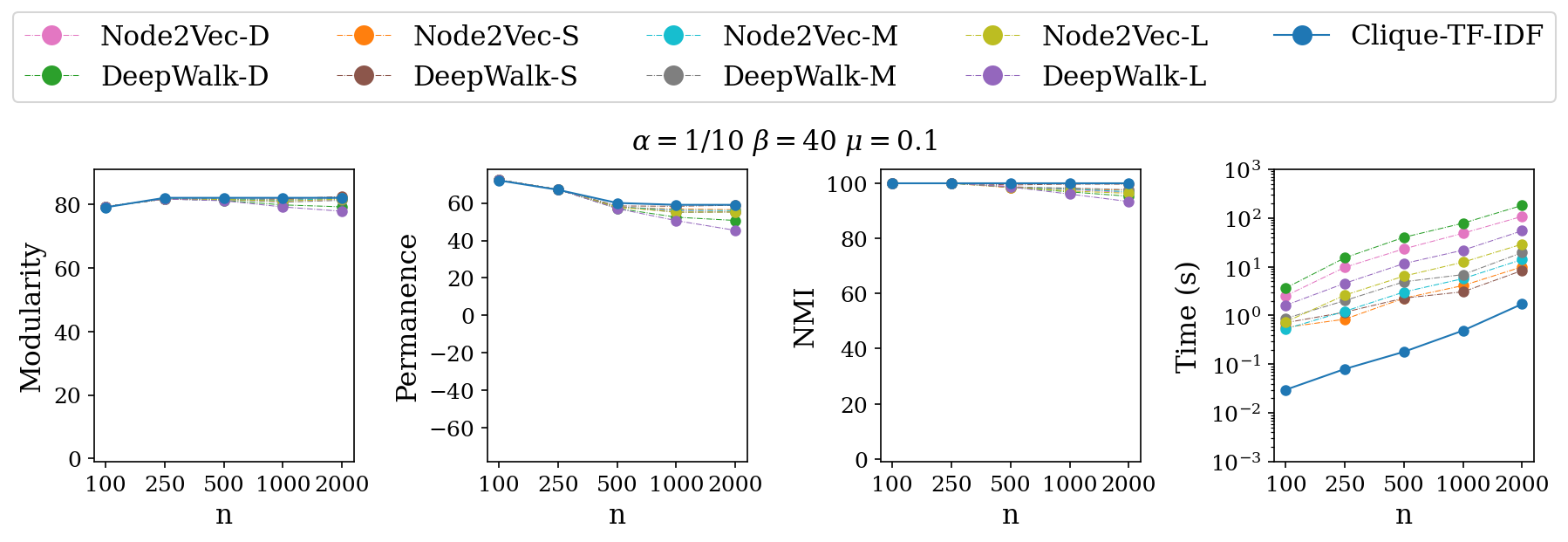} \\
~~~~~{(a)}\hspace{\foursep}{(b)}\hspace{\foursep}{(c)}\hspace{\foursep}{(d)} \\
\includegraphics[width=1\linewidth]{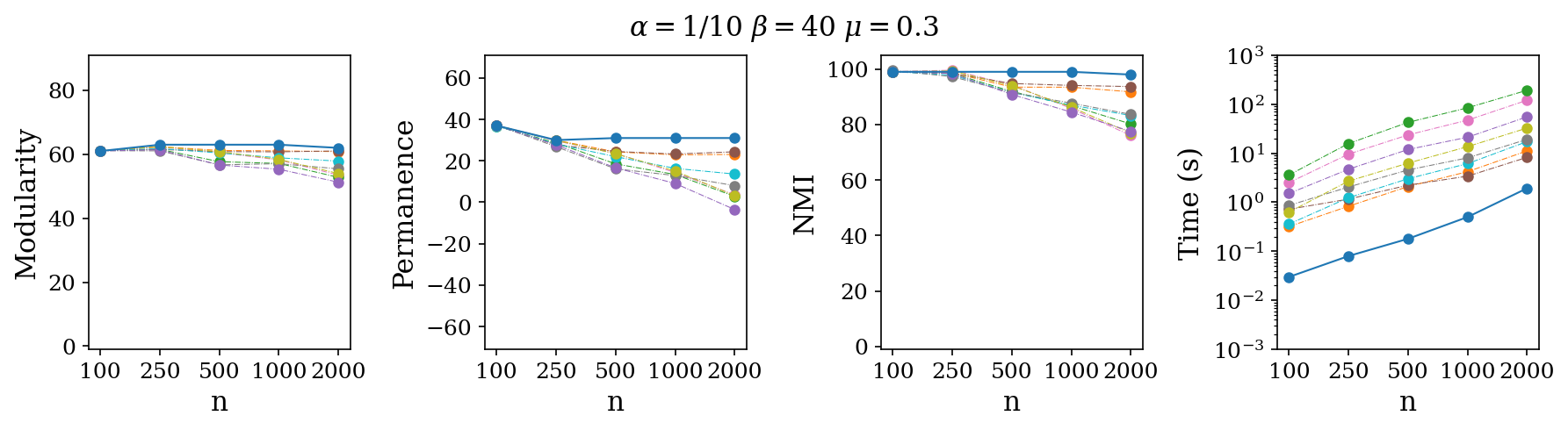} \\
~~~~~{(e)}\hspace{\foursep}{(f)}\hspace{\foursep}{(g)}\hspace{\foursep}{(h)} \\
\caption{Evaluation metrics on synthetic networks when $\alpha=1/10$ and $\beta=40$, varying $n$, for $\mu \in \{0.1,0.3\}$ (rows).} 
\label{fi:embedding-synthetic-varying-n}
\end{figure*}

\smallskip \noindent{\bf Varying the number of vertices.}
\cref{fi:embedding-synthetic-varying-n} shows the evaluation with respect to different values of $n$, choosing $\alpha=1/10$ and $\beta=40$. When $\mu=0.1$, our approach achieves the highest results in terms of quality, and in particular, {\em Clique-TF-IDF} obtains values of modularity that are independent of the value of $n$, while all considered variants of Node2Vec and DeepWalk tend to decrease their quality as $n$ increases (see \Cref{fi:embedding-synthetic-varying-n}(a)-(c)). This phenomenon is amplified when the network is noisier, (i.e., when the mixing parameter increases), with Node2Vec and DeepWalk obtaining results far below those obtained by our algorithm, as it can be seen in \Cref{fi:embedding-synthetic-varying-n}(e)-(g).

Furthermore, when analyzing the running time, \cref{fi:embedding-synthetic-varying-n}(d) and \cref{fi:embedding-synthetic-varying-n}(h) show that the embedding phase of {\em Clique-TF-IDF} is at least one order of magnitude faster with respect to all the other embedding algorithms considered.

\begin{figure*}[htb]
\centering
\includegraphics[width=1\linewidth]{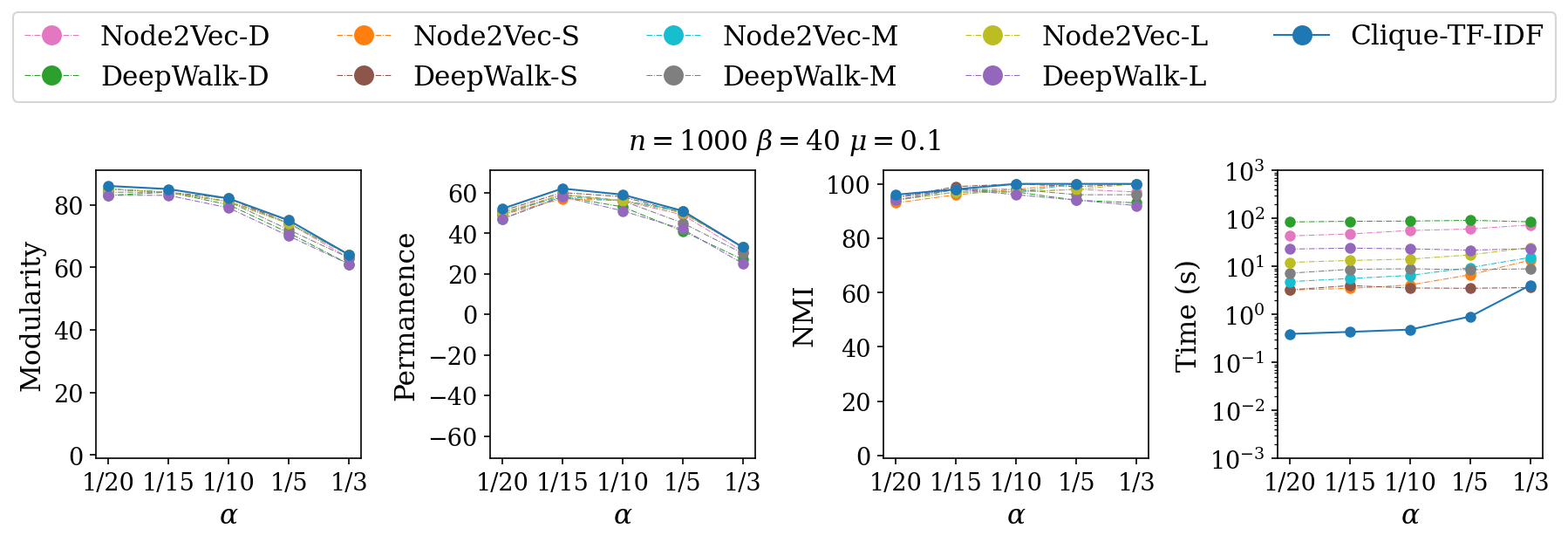} \\
~~~~~{(a)}\hspace{\foursep}{(b)}\hspace{\foursep}{(c)}\hspace{\foursep}{(d)} \\
 \includegraphics[width=1\linewidth]{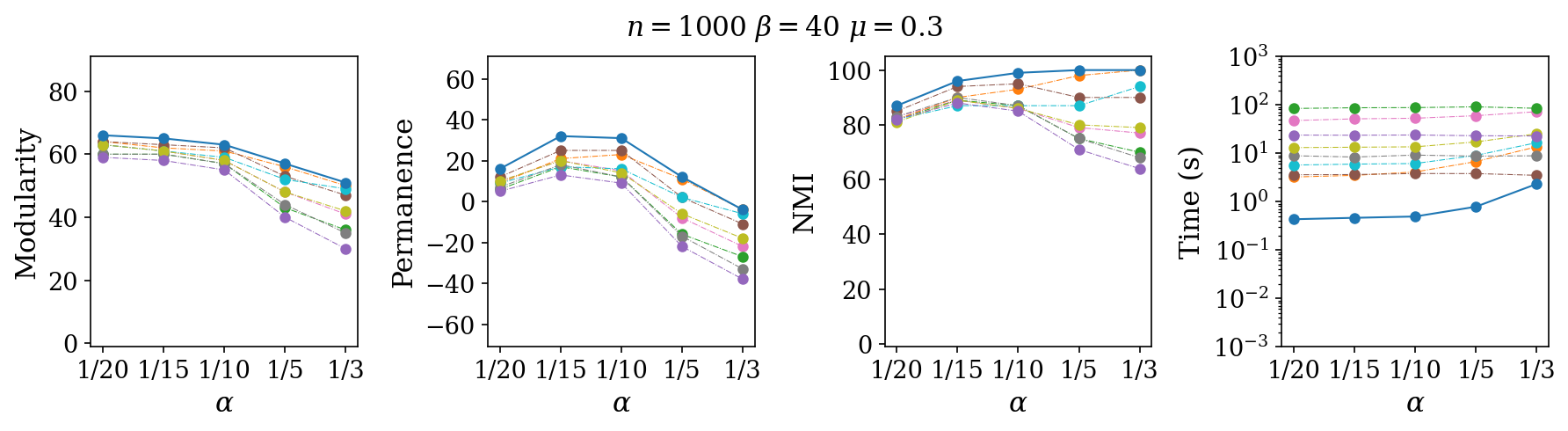} \\
~~~~~{(e)}\hspace{\foursep}{(f)}\hspace{\foursep}{(g)}\hspace{\foursep}{(h)} \\
\caption{Evaluation metrics on synthetic networks when $n=1000$ and $\beta=40$, varying $\alpha$, for $\mu \in \{0.1,0.3\}$ (rows).}  
\label{fi:embedding-synthetic-varying-alpha}
\end{figure*}
\smallskip \noindent{\bf Varying maximum vertex degree.} In the end, we analyze how the scale-freeness of the network affects the embedding. We choose $n$ and $\beta$ to be 1000 and 40 respectively. When $\mu=0.1$, {\em Clique-TF-IDF} obtains values of modularity and permanence that decrease as $\alpha$ increases, but having results that are always better than other approaches (see \Cref{fi:embedding-synthetic-varying-alpha}(a)-(b)). This reduction is driven by the structure of the network, as discussed in \cref{sss:synthetic}. 

Furthermore, when comparing to the underlying ground truth, our approach is able to discover most of the communities when $\alpha < 1/10$, while for higher values we are able to find the exact ground truth structure when on the contrary Node2Vec and DeepWalk obtain lower results, as it can be seen in \cref{fi:embedding-synthetic-varying-alpha}(c).

Once again, our algorithm performs better in terms of modularity, permanence and NMI for all the values of $\alpha$ when the mixing parameter $\mu$ increases to 0.3, while other approaches lead to results that diverge from our curves as $\alpha$ increases (see \Cref{fi:embedding-synthetic-varying-alpha}(e)-(g)).

When considering efficiency, we observe in \cref{fi:embedding-synthetic-varying-alpha}(d) and \cref{fi:embedding-synthetic-varying-alpha}(h) that {\em Clique-TF-IDF} exhibits a stronger dependency on the running time with respect to the value of $\alpha$ due to the larger number of maximal cliques (see \cref{fi:clique-distribution-synthetic-varying-alpha-and-mu}). However, it consistently calculates embeddings more efficiently than Node2Vec and DeepWalk.

\section{Conclusions and future work}\label{se:conclusions}

We introduced {\em Clique-TF-IDF}, a novel framework for graph partitioning that leverages dense graph-theoretic subgraphs, modeled as maximal cliques, to describe each vertex in terms of the cliques it is part of. 
In addition, we justified the choice of computing a vertex positional embedding based on maximal cliques.
{\em Clique-TF-IDF} is adaptable to various algorithms, accommodating scenarios where the number $k$ of communities is either given in advance or unknown.

The high computation time is a trade-off for the quality of partitions produced by {\em Clique-TF-IDF}. Our experimental results show that, when the number $k$ of partition blocks is predetermined, {\em Clique-TF-IDF} outperforms the most effective algorithms in the literature in terms of effectiveness. Even in cases where $k$ is not known a priori, the proposed approach remains promising, producing partitions with quality comparable to that of state-of-the-art competitors.

Our approach requires, as a preliminary step, computing the maximal cliques of a graph: this might appear to be prohibitive as, in theory, the number of maximal cliques could be exponential in the number of vertices. However, a large research literature guarantees that maximal cliques can be efficiently computed on sparse real-world graphs, also leveraging distributed approaches. Hence, in practical cases, the efficiency of {\em Clique-TF-IDF} is limited not by the computation of cliques itself, but rather by their number (as observed in \cref{se:experiments-real}). Indeed, in our current implementation, we handle matrices whose dimensions are comparable with the number of cliques, which is largely independent of the number of vertices of the graph and is usually limited for social networks.
In our experiments, on a small platform with limited main memory size (see \cref{sse:setup}), we could handle graphs up to approximately ten thousand vertices, one hundred thousand edges and maximal cliques in less than one minute. 

In scenarios where computational time is not a primary concern and the input graph contains up to several hundred thousand cliques, the approach proposed in this paper offers high-quality solutions that rival -- or even surpass -- traditional algorithms.
One of the key findings of our study is that our approach, which combines AI techniques with pioneering combinatorial algorithms, produced results comparable to state-of-the-art algorithms developed over decades of research for complex problems.
In addition to applying AI to other intractable combinatorial problems where substructures can be efficiently enumerated, we suggest three promising research directions for future work:
\begin{itemize}
\item The high running time of the approach, especially when the input network has a large number of maximal cliques, could be reduced by applying a clique sampling technique similar to that proposed in ~\cite{DBLP:conf/kdd/WangCF13,DBLP:conf/icde/Li0CZL0021}. The overall approach could also benefit of a distributed implementation, leveraging distributed maximal cliques enumeration algorithms~\cite{DBLP:journals/tods/ChengKFYZ11,DBLP:conf/kdd/ChengZKC12,DBLP:conf/edbt/ConteVMPT16,xu2014distributed}.
\item Replacing maximal cliques, which serve as the starting point for the embedding, with other graph-theoretic dense substructures, such as trusses~\cite{JJ12}, $k$-plexes~\cite{cfpt-snde2-19}, fixed-size cliques~\cite{CoppaFG19,FinocchiFF15}, quasi-cliques~\cite{DBLP:journals/isci/MeloRR22,DBLP:journals/isci/PengWWW21}, or $k$-diamonds~\cite{FGS23}, might improve the quality of produced partitions.
\item Last but not least, a similar strategy could be effectively applied to the computation of structural clusterings~\cite{DBLP:conf/kdd/RibeiroSF17,DBLP:conf/iclr/Srinivasan020,DBLP:conf/sdm/ZhuLHK21}.
\end{itemize}



\bibliography{biblio}

@article{DBLP:journals/csr/DEliaFP25,
  author       = {Marco D'Elia and
                  Irene Finocchi and
                  Maurizio Patrignani},
  title        = {Maximal cliques summarization: Principles, problem classification,
                  and algorithmic approaches},
  journal      = {Comput. Sci. Rev.},
  volume       = {58},
  pages        = {100784},
  year         = {2025},
  urlREMOVED          = {https://doi.org/10.1016/j.cosrev.2025.100784},
  doi          = {10.1016/J.COSREV.2025.100784},
  bibsource    = {dblp computer science bibliography, https://dblp.org}
}

@article{DBLP:journals/isci/DrazdilovaPPS24,
  author       = {Pavla Dr{\'{a}}zdilov{\'{a}} and
                  Petr Prokop and
                  Jan Platos and
                  V{\'{a}}clav Sn{\'{a}}sel},
  title        = {A hierarchical overlapping community detection method based on closed
                  trail distance and maximal cliques},
  journal      = {Inf. Sci.},
  volume       = {662},
  pages        = {120271},
  year         = {2024},
  urlREMOVED          = {https://doi.org/10.1016/j.ins.2024.120271},
  doi          = {10.1016/J.INS.2024.120271},
  bibsource    = {dblp computer science bibliography, https://dblp.org}
}

@article{DBLP:journals/isci/GlariaHLNS21,
  author       = {Felipe Glaria and
                  Cecilia Hern{\'{a}}ndez and
                  Susana Ladra and
                  Gonzalo Navarro and
                  Lilian Salinas},
  title        = {Compact structure for sparse undirected graphs based on a clique graph
                  partition},
  journal      = {Inf. Sci.},
  volume       = {544},
  pages        = {485--499},
  year         = {2021},
  urlREMOVED          = {https://doi.org/10.1016/j.ins.2020.09.010},
  doi          = {10.1016/J.INS.2020.09.010},
  bibsource    = {dblp computer science bibliography, https://dblp.org}
}

@article{DBLP:journals/isci/PengWWW21,
  author       = {Bo Peng and
                  Lifan Wu and
                  Yang Wang and
                  Qinghua Wu},
  title        = {Solving maximum quasi-clique problem by a hybrid artificial bee colony
                  approach},
  journal      = {Inf. Sci.},
  volume       = {578},
  pages        = {214--235},
  year         = {2021},
  urlREMOVED          = {https://doi.org/10.1016/j.ins.2021.06.094},
  doi          = {10.1016/J.INS.2021.06.094},
  bibsource    = {dblp computer science bibliography, https://dblp.org}
}

@article{DBLP:journals/isci/MeloRR22,
  author       = {Rafael A. Melo and
                  Celso C. Ribeiro and
                  Jos{\'{e}} A. Riveaux},
  title        = {The minimum quasi-clique partitioning problem: Complexity, formulations,
                  and a computational study},
  journal      = {Inf. Sci.},
  volume       = {612},
  pages        = {655--674},
  year         = {2022},
  urlREMOVED          = {https://doi.org/10.1016/j.ins.2022.08.073},
  doi          = {10.1016/J.INS.2022.08.073},
  bibsource    = {dblp computer science bibliography, https://dblp.org}
}

@article{conte-bipartite,
  author       = {Alessio Conte and
                  Roberto Grossi and
                  Andrea Marino and
                  Takeaki Uno and
                  Luca Versari},
  title        = {Proximity Search for Maximal Subgraph Enumeration},
  journal      = {{SIAM} J. Comput.},
  volume       = {51},
  number       = {5},
  pages        = {1580--1625},
  year         = {2022},
  doi         = {10.1137/20M1375048}
}

@article{CoppaFG19,
  author       = {Emilio Coppa and
                  Irene Finocchi and
                  Renan Leon Garcia},
  title        = {Counting cliques in parallel without a cluster: Engineering a fork/join
                  algorithm for shared-memory platforms},
  journal      = {Inf. Sci.},
  volume       = {496},
  pages        = {553--571},
  year         = {2019},
  doi          = {10.1016/J.INS.2018.07.018},
  
}

@article{JJ12,
author = {Wang, Jia and Cheng, James},
title = {Truss Decomposition in Massive Networks},
year = {2012},
publisher = {VLDB Endowment},
volume = {5},
number = {9},
journal = {Proc. VLDB Endow.},
pages = {812–823},
doi          = {10.14778/2311906.2311909},
}

@inproceedings{cfpt-snde2-19,	author="Alessio Conte and Donatella Firmani and Maurizio Patrignani and Riccardo Torlone",
	title="Shared-Nothing Distributed Enumeration of 2-Plexes",
booktitleREMOVED="28th ACM International Conference on Information and Knowledge Management (CIKM 2019)",
booktitle = "CIKM 2019",
	year="2019",
	editorOPTIONAL="Peng Cui and Elke Rundensteiner and David Carmel and Qi He and {Jeffrey Xu} Yu",	publisher="ACM", address="New York",
	pages="2469-2472",
	isbn="978-1-4503-6976-3",
	doi="https://doi.org/10.1145/3357384.3358083"
}

@article{FGS23,
    author = {Finocchi, Irene and Garcia, Renan Leon and Sinaimeri, Blerina},
    title = "{From stars to diamonds: Counting and listing almost complete subgraphs in large networks}",
    journal = {The Computer Journal,},
    volume = {67(6)},
    year = {2024},
    doi          = {10.1093/COMJNL/BXAD129},
    issn = {0010-4620},
}

@article{FinocchiFF15,
  author       = {Irene Finocchi and
                  Marco Finocchi and
                  Emanuele G. Fusco},
  title        = {Clique Counting in {MapReduce}: Algorithms and Experiments},
  journal      = {{ACM} J. Exp. Algorithmics},
  volume       = {20},
  pages        = {1.7:1--1.7:20},
  year         = {2015},
  url_removed          = {https://doi.org/10.1145/2794080},
  doi          = {10.1145/2794080},
  biburl_removed       = {https://dblp.org/rec/journals/jea/FinocchiFF15.bib}
}

@inproceedings{DBLP:conf/coco/Karp72,
  author    = {Richard M. Karp},
  editor    = {Raymond E. Miller and
               James W. Thatcher},
  title     = {Reducibility Among Combinatorial Problems},
  booktitleREMOVED = {Proceedings of a symposium on the Complexity of Computer Computations,
               held March 20-22, 1972, at the {IBM} Thomas J. Watson Research Center,
               Yorktown Heights, New York, {USA}},
  booktitle = {Proc. Symp. Compl. Comp. Computat.},
  seriesOPTIONAL    = {The {IBM} Research Symposia Series},
  pages     = {85--103},
  publisher = {Plenum Press}, 
  address = {New York},
  year      = {1972},
  urlREMOVED       = {https://doi.org/10.1007/978-1-4684-2001-2\_9},
  doi       = {10.1007/978-1-4684-2001-2\_9}
}

@article{DBLP:journals/cacm/BronK73,
  author    = {Coenraad Bron and
               Joep Kerbosch},
  title     = {Finding All Cliques of an Undirected Graph (Algorithm 457)},
  journal   = {Commun. {ACM}},
  volume    = {16},
  number    = {9},
  pages     = {575--576},
  year      = {1973},
  bibsource = {dblp computer science bibliography, https://dblp.org},
  doi = {10.1145/362342.362367},
}

@inproceedings{DBLP:conf/cocoon/TomitaTT04,
  author    = {Etsuji Tomita and
               Akira Tanaka and
               Haruhisa Takahashi},
  editor    = {Kyung{-}Yong Chwa and
               J. Ian Munro},
  title     = {The Worst-Case Time Complexity for Generating All Maximal Cliques},
  booktitleREMOVED = {Computing and Combinatorics, 10th Annual International Conference,
               {COCOON} 2004, Jeju Island, Korea, August 17-20, 2004, Proceedings},
booktitle = {{COCOON} 2004},
  seriesREMOVE    = {Lecture Notes in Computer Science},
  volume    = {3106},
  pages     = {161--170},
  publisher = {Springer},
  address = {Germany},
  year      = {2004},
  urlREMOVED       = {https://doi.org/10.1007/978-3-540-27798-9\_19},
  doi       = {10.1007/978-3-540-27798-9\_19}
}

@article{DBLP:journals/tcs/CazalsK08,
  author    = {Fr{\'{e}}d{\'{e}}ric Cazals and
               Chinmay Karande},
  title     = {A note on the problem of reporting maximal cliques},
  journal   = {Theor. Comput. Sci.},
  volume    = {407},
  number    = {1-3},
  pages     = {564--568},
  year      = {2008},
  urlREMOVED       = {https://doi.org/10.1016/j.tcs.2008.05.010},
  doi       = {10.1016/j.tcs.2008.05.010}
}

@article{DBLP:journals/tcs/Koch01,
  author    = {Ina Koch},
  title     = {Enumerating all connected maximal common subgraphs in two graphs},
  journal   = {Theor. Comput. Sci.},
  volume    = {250},
  number    = {1-2},
  pages     = {1--30},
  year      = {2001},
  urlREMOVED       = {https://doi.org/10.1016/S0304-3975(00)00286-3},
  doi       = {10.1016/S0304-3975(00)00286-3}
}

@article{DBLP:conf/wea/EppsteinS11,
  author       = {David Eppstein and
                  Maarten L{\"{o}}ffler and
                  Darren Strash},
  title        = {Listing All Maximal Cliques in Large Sparse Real-World Graphs},
  journal      = {{ACM} J. Exp. Algorithmics},
  volume       = {18},
  year         = {2013},
  doi          = {10.1145/2543629},
  bibsource    = {dblp computer science bibliography, https://dblp.org}
}

@inproceedings{DBLP:conf/edbt/ConteVMPT16,
  author    = {Alessio Conte and
               Roberto {De Virgilio} and
               Antonio Maccioni and
               Maurizio Patrignani and
               Riccardo Torlone},
  editorOPTIONAL    = {Evaggelia Pitoura and
               Sofian Maabout and
               Georgia Koutrika and
               Am{\'{e}}lie Marian and
               Letizia Tanca and
               Ioana Manolescu and
               Kostas Stefanidis},
  title     = {Finding All Maximal Cliques in Very Large Social Networks},
  booktitleOPTIONAL = {Proceedings of the 19th International Conference on Extending Database
               Technology, {EDBT} 2016, Bordeaux, France, March 15-16, 2016, Bordeaux,
               France, March 15-16, 2016},
  booktitle = {{EDBT} 2016},
  pages     = {173--184},
  publisher = {OpenProceedings.org},
  address = {Konstanz, Germany},
  year      = {2016},
  urlOPTIONAL       = {https://doi.org/10.5441/002/edbt.2016.18},
  doi       = {10.5441/002/edbt.2016.18},
}

@article{DBLP:journals/tods/ChengKFYZ11,
  author    = {James Cheng and
               Yiping Ke and
               Ada Wai-Chee Fu and
               Jeffrey Xu Yu and
               Linhong Zhu},
  title     = {Finding maximal cliques in massive networks},
  journal   = {ACM Trans. Database Syst.},
  doi          = {10.1145/2043652.2043654},
  volume    = {36},
  number    = {4},
  year      = {2011},
  pages     = {21}
}

@inproceedings{DBLP:conf/kdd/ChengZKC12,
  author    = {James Cheng and
               Linhong Zhu and
               Yiping Ke and
               Shumo Chu},
  title     = {Fast algorithms for maximal clique enumeration with limited
               memory},
  booktitle = {KDD},
  year      = {2012},
  pages     = {1240-1248},
  doi          = {10.1145/2339530.2339724},
}

@inproceedings{xu2014distributed,
author       = {Yanyan Xu and
                  James Cheng and
                  Ada Wai{-}Chee Fu and
                  Yingyi Bu},
  title        = {Distributed Maximal Clique Computation},
  booktitle    = {2014 {IEEE} International Congress on Big Data, Anchorage, AK, USA,
                  June 27 - July 2, 2014},
  pages        = {160--167},
  publisher    = {{IEEE} Computer Society},
  year         = {2014},
  urlREMOVED          = {https://doi.org/10.1109/BigData.Congress.2014.31},
  doi          = {10.1109/BIGDATA.CONGRESS.2014.31},
  bibsource    = {dblp computer science bibliography, https://dblp.org}
}

@Incollection{Pattillo2012,
  author = "Pattillo, Jeffrey and Youssef, Nataly and Butenko, Sergiy",
  editor = "Thai, My T. and Pardalos, Panos M.",
  title="Clique Relaxation Models in Social Network Analysis",
  bookTitle="Handbook of Optimization in Complex Networks: Communication and Social Networks",
  year="2012",
  publisher="Springer New York",
  address="New York, NY",
  pages="143--162",
  isbn="978-1-4614-0857-4",
  doi="10.1007/978-1-4614-0857-4\_5"
}

@inproceedings{DBLP:conf/wg/BrandesDGGHNW07,
  author    = {Ulrik Brandes and
               Daniel Delling and
               Marco Gaertler and
               Robert G{\"{o}}rke and
               Martin Hoefer and
               Zoran Nikoloski and
               Dorothea Wagner},
  editorOPTIONAL    = {Andreas Brandst{\"{a}}dt and
               Dieter Kratsch and
               Haiko M{\"{u}}ller},
  title     = {On Finding Graph Clusterings with Maximum Modularity},
  booktitle = {{WG} 2007},
  booktitleOPTIONAL = {Graph-Theoretic Concepts in Computer Science, 33rd International Workshop,
               {WG} 2007, Dornburg, Germany, June 21-23, 2007. Revised Papers},
  seriesOPTIONAL    = {Lecture Notes in Computer Science},
  volume    = {4769},
  pages     = {121--132},
  publisher = {Springer},
  address = {Germany},
  year      = {2007},
  urlREMOVED       = {https://doi.org/10.1007/978-3-540-74839-7\_12},
  doi       = {10.1007/978-3-540-74839-7\_12}
}

@article{Girvan_2002,
	doi = {10.1073/pnas.122653799},
	urlREMOVED = {https://doi.org/10.1073%2Fpnas.122653799},
	year = 2002,
	month = {jun},
	publisher = {Proceedings of the National Academy of Sciences},
	volume = {99},
	number = {12},
	pages = {7821--7826},
	author = {M. Girvan and M. E. J. Newman},
	title = {Community structure in social and biological networks},
	journal = {Proceedings of the National Academy of Sciences}
}

@article{PhysRevE.70.066111,
  title = {Finding community structure in very large networks},
  author = {Clauset, Aaron and Newman, M. E. J. and Moore, Cristopher},
  journal = {Phys. Rev. E},
  volume = {70},
  issue = {6},
  pages = {066111},
  numpages = {6},
  year = {2004},
  month = {Dec},
  publisher = {American Physical Society},
  doi = {10.1103/PhysRevE.70.066111},
  urlOPTIONAL = {https://link.aps.org/doi/10.1103/PhysRevE.70.066111}
}

@article{Blondel_2008,
	doi = {10.1088/1742-5468/2008/10/p10008},
	urlREMOVED = {https://doi.org/10.1088/1742-5468/2008/10/p10008},
	year = 2008,
	month = {oct},
	publisher = {{IOP} Publishing},
	volume = {2008},
	number = {10},
	pages = {P10008},
	author = {Vincent D Blondel and Jean-Loup Guillaume and Renaud Lambiotte and Etienne Lefebvre},
	title = {Fast unfolding of communities in large networks},
	journal = {Journal of Statistical Mechanics: Theory and Experiment}
}

@inproceedings{10.1145/2566486.2568010,
  author = {Prat-P\'{e}rez, Arnau and Dominguez-Sal, David and Larriba-Pey, Josep-Lluis},
  title = {High Quality, Scalable and Parallel Community Detection for Large Real Graphs},
  year = {2014},
  isbnREMOVED = {9781450327442},
  publisher = {Association for Computing Machinery},
  address = {New York, NY, USA},
  urlREMOVED = {https://doi.org/10.1145/2566486.2568010},
  doi = {10.1145/2566486.2568010},
  booktitleREMOVED = {Proceedings of the 23rd International Conference on World Wide Web},
  booktitle = {Proc. WWW 2014},
  pages = {225–236},
  numpages = {12},
  location = {Seoul, Korea},
  seriesREMOVED = {WWW '14}
}

@article{Cordasco2010CommunityDV,
  title={Community detection via semi-synchronous label propagation algorithms},
  author={Gennaro Cordasco and Luisa Gargano},
  journal={2010 IEEE International Workshop on: Business Applications of Social Network Analysis (BASNA)},
  year={2010},
  pages={1-8}
}

@misc{https://doi.org/10.48550/arxiv.physics/0512106,
  doi = {10.48550/ARXIV.PHYSICS/0512106},
  urlOPTIONAL = {https://arxiv.org/abs/physics/0512106},
  author = {Pons, Pascal and Latapy, Matthieu},
  title = {Computing communities in large networks using random walks (long version)},
  publisher = {arXiv},
  year = {2005},
  copyright = {Assumed arXiv.org perpetual, non-exclusive license to distribute this article for submissions made before January 2004}
}

@article{Reichardt_2006,
	doi = {10.1103/physreve.74.016110},
	urlRemoved= {https://doi.org/10.1103\%2Fphysreve.74.016110},
	year = 2006,
	month = {jul},
	publisher = {American Physical Society ({APS})},
	volume = {74},
	number = {1},
	author = {Jörg Reichardt and Stefan Bornholdt},
	title = {Statistical mechanics of community detection},
	journal = {Physical Review E}
}

@inproceedings{10.5555/2951659.2951772,
author = {Pizzuti, Clara},
  title = {GA-Net: A Genetic Algorithm for Community Detection in Social Networks},
  year = {2008},
  isbn = {9783540876991},
  publisher = {Springer-Verlag},
  OPTaddress = {Berlin, Heidelberg},
  booktitle = {Proc. of the 10th Int. Conf. on Parallel Problem Solving from Nature (PPSN X), Vol. 5199},
  pages = {1081–1090},
  doi          = {10.1007/978-3-540-87700-4\_107},
  numpages = {10}
}

@article{doi:10.1073/pnas.0706851105,
author = {Martin Rosvall  and Carl T. Bergstrom },
title = {Maps of random walks on complex networks reveal community structure},
journal = {Proceedings of the National Academy of Sciences},
volume = {105},
number = {4},
pages = {1118-1123},
year = {2008},
doi = {10.1073/pnas.0706851105},
URLREMOVED = {https://www.pnas.org/doi/abs/10.1073/pnas.0706851105},
eprintREMOVED = {https://www.pnas.org/doi/pdf/10.1073/pnas.0706851105}}

@article{Leiden2019,
  author       = {Vincent A. Traag and
                  Ludo Waltman and
                  Nees Jan van Eck},
  title        = {From Louvain to Leiden: guaranteeing well-connected communities},
  journal      = {Scientific Reports},
  volume       = {9},
  year         = {2019},
}

@book{manning2008introduction,
  address = {Cambridge, UK},
  author = {Manning, Christopher D. and Raghavan, Prabhakar and Schütze, Hinrich},
  isbn = {978-0-521-86571-5},
  publisher = {Cambridge University Press},
  title = {Introduction to Information Retrieval},
  urlRemoved = {http://nlp.stanford.edu/IR-book/information-retrieval-book.html},
  year = 2008
}

@book{10.5555/3208440,
author = {Tan, Pang-Ning and Steinbach, Michael and Karpatne, Anuj and Kumar, Vipin},
title = {Introduction to Data Mining (2nd Edition)},
year = {2018},
isbn = {0133128903},
publisher = {Pearson},
address = {~\hspace{-0.5em}},
edition = {2nd}
}

@inproceedings{DBLP:conf/iclr/Srinivasan020,
  author       = {Balasubramaniam Srinivasan and
                  Bruno Ribeiro},
  title        = {On the Equivalence between Positional Node Embeddings and Structural
                  Graph Representations},
  booktitleOPTIONAL    = {8th International Conference on Learning Representations, {ICLR} 2020,
                  Addis Ababa, Ethiopia, April 26-30, 2020},
  booktitle    = {{ICLR} 2020},
  publisher    = {OpenReview.net},
  year         = {2020},
  address = {~\hspace{-0.5em}},
  urlREMOVED          = {https://openreview.net/forum?id=SJxzFySKwH}
}

@misc{snapnets,
  author       = {Jure Leskovec and Andrej Krevl},
  title        = {{SNAP Datasets}: {Stanford} Large Network Dataset Collection},
  howpublished = {\url{http://snap.stanford.edu/data}},
  month        = jun,
  year         = 2014
}

@inproceedings{DBLP:conf/nips/DevvritSD022,
  author       = {Fnu Devvrit and
                  Aditya Sinha and
                  Inderjit S. Dhillon and
                  Prateek Jain},
  title        = {{S3GC:} Scalable Self-Supervised Graph Clustering},
  booktitle    = {NeurIPS},
  year         = {2022},
  urlREMOVED         = {http://papers.nips.cc/paper\_files/paper/2022/hash/15972a9575e0f03bf82f00aebeb40774-Abstract-Conference.html},
  bibsource    = {dblp computer science bibliography, https://dblp.org}
}

@inproceedings{DBLP:conf/ida/CombeLGE15,
  author       = {David Combe and
                  Christine Largeron and
                  Mathias G{\'{e}}ry and
                  El{\"{o}}d Egyed{-}Zsigmond},
  editor       = {{\'{E}}lisa Fromont and
                  Tijl De Bie and
                  Matthijs van Leeuwen},
  title        = {I-Louvain: An Attributed Graph Clustering Method},
  booktitleREMOVED    = {Advances in Intelligent Data Analysis {XIV} - 14th International Symposium,
                  {IDA} 2015, Saint Etienne, France, October 22-24, 2015, Proceedings},
  booktitle    = {{IDA} 2015},
  series       = {Lecture Notes in Computer Science},
  volume       = {9385},
  pages        = {181--192},
  publisher    = {Springer},
  address = {Germany},
  year         = {2015},
  urlREMOVED          = {https://doi.org/10.1007/978-3-319-24465-5\_16},
  doi         = {10.1007/978-3-319-24465-5\_16},
  bibsource    = {dblp computer science bibliography, https://dblp.org}
}

@inproceedings{DBLP:conf/kdd/GroverL16,
  author       = {Aditya Grover and
                  Jure Leskovec},
  editor       = {Balaji Krishnapuram and
                  Mohak Shah and
                  Alexander J. Smola and
                  Charu C. Aggarwal and
                  Dou Shen and
                  Rajeev Rastogi},
  title        = {node2vec: Scalable Feature Learning for Networks},
  booktitleREMOVED    = {Proceedings of the 22nd {ACM} {SIGKDD} International Conference on
                  Knowledge Discovery and Data Mining, San Francisco, CA, USA, August
                  13-17, 2016},
  booktitle    = {{ACM} {SIGKDD} 2016},
  pages        = {855--864},
  publisher    = {{ACM}},
  address = {USA},
  year         = {2016},
  urlREMOVED          = {https://doi.org/10.1145/2939672.2939754},
  doi         = {10.1145/2939672.2939754}
}

@inproceedings{DBLP:conf/kdd/PerozziAS14,
  author       = {Bryan Perozzi and
                  Rami Al{-}Rfou and
                  Steven Skiena},
  editor       = {Sofus A. Macskassy and
                  Claudia Perlich and
                  Jure Leskovec and
                  Wei Wang and
                  Rayid Ghani},
  title        = {DeepWalk: online learning of social representations},
  booktitle    = {The 20th {ACM} {SIGKDD} International Conference on Knowledge Discovery
                  and Data Mining, {KDD} '14, New York, NY, {USA} - August 24 - 27,
                  2014},
  pages        = {701--710},
  publisher    = {{ACM}},
  address = {USA},
  year         = {2014},
  urlREMOVED          = {https://doi.org/10.1145/2623330.2623732},
  doi         = {10.1145/2623330.2623732},
  bibsource    = {dblp computer science bibliography, https://dblp.org}
}

@inproceedings{DBLP:conf/nips/LattanziMVWZ21,
  author       = {Silvio Lattanzi and
                  Benjamin Moseley and
                  Sergei Vassilvitskii and
                  Yuyan Wang and
                  Rudy Zhou},
  editor       = {Marc'Aurelio Ranzato and
                  Alina Beygelzimer and
                  Yann N. Dauphin and
                  Percy Liang and
                  Jennifer Wortman Vaughan},
  title        = {Robust Online Correlation Clustering},
  booktitleREMOVED    = {Advances in Neural Information Processing Systems 34: Annual Conference
                  on Neural Information Processing Systems 2021, NeurIPS 2021, December
                  6-14, 2021, virtual},
  booktitle    = {NeurIPS 2021},
  pages        = {4688--4698},
  year         = {2021},
  urlREMOVED          = {https://proceedings.neurips.cc/paper/2021/hash/250dd56814ad7c50971ee4020519c6f5-Abstract.html}
}

@article{DBLP:journals/ml/BansalBC04,
  author       = {Nikhil Bansal and
                  Avrim Blum and
                  Shuchi Chawla},
  title        = {Correlation Clustering},
  journal      = {Mach. Learn.},
  volume       = {56},
  number       = {1-3},
  pages        = {89--113},
  year         = {2004},
  urlREMOVED          = {https://doi.org/10.1023/B:MACH.0000033116.57574.95},
  doi         = {10.1023/B:MACH.0000033116.57574.95},
  bibsource    = {dblp computer science bibliography, https://dblp.org}
}

@inproceedings{DBLP:conf/esa/SahaS19,
  author       = {Barna Saha and Sanjay Subramanian},
  editor       = {Michael A. Bender and
                  Ola Svensson and
                  Grzegorz Herman},
  title        = {Correlation Clustering with Same-Cluster Queries Bounded by Optimal
                  Cost},
  booktitleREMOVED    = {27th Annual European Symposium on Algorithms, {ESA} 2019, September
                  9-11, 2019, Munich/Garching, Germany},
  booktitle    = {{ESA} 2019},
  series       = {LIPIcs},
  volume       = {144},
  pages        = {81:1--81:17},
  publisher    = {Schloss Dagstuhl - Leibniz-Zentrum f{\"{u}}r Informatik},
  address = {Germany},
  year         = {2019},
  urlREMOVED          = {https://doi.org/10.4230/LIPIcs.ESA.2019.81},
  doi         = {10.4230/LIPIcs.ESA.2019.81},
  bibsource    = {dblp computer science bibliography, https://dblp.org}
}

@inproceedings{DBLP:conf/kdd/RibeiroSF17,
  author       = {Leonardo Filipe Rodrigues Ribeiro and
                  Pedro H. P. Saverese and
                  Daniel R. Figueiredo},
  title        = {\emph{struc2vec}: Learning Node Representations from Structural Identity},
  booktitleREMOVED    = {Proceedings of the 23rd {ACM} {SIGKDD} International Conference on
                  Knowledge Discovery and Data Mining, Halifax, NS, Canada, August 13
                  - 17, 2017},
  booktitle    = {{ACM} {SIGKDD} 2017},
  pages        = {385--394},
  publisher    = {{ACM}},
  address = {USA},
  year         = {2017},
  urlREMOVED          = {https://doi.org/10.1145/3097983.3098061},
  doi         = {10.1145/3097983.3098061},
  bibsource    = {dblp computer science bibliography, https://dblp.org}
}

@article{DBLP:journals/nn/ChenYHLPWZ23,
  author       = {Yangrui Chen and
                  Jiaxuan You and
                  Jun He and
                  Yuan Lin and
                  Yanghua Peng and
                  Chuan Wu and
                  Yibo Zhu},
  title        = {{SP-GNN:} Learning structure and position information from graphs},
  journal      = {Neural Networks},
  volume       = {161},
  pages        = {505--514},
  year         = {2023},
  urlREMOVED          = {https://doi.org/10.1016/j.neunet.2023.01.051},
  doi         = {10.1016/j.neunet.2023.01.051},
  bibsource    = {dblp computer science bibliography, https://dblp.org}
}

@inproceedings{DBLP:conf/sdm/ZhuLHK21,
  author       = {Jing Zhu and
                  Xingyu Lu and
                  Mark Heimann and
                  Danai Koutra},
  editor       = {Carlotta Demeniconi and
                  Ian Davidson},
  title        = {Node Proximity Is All You Need: Unified Structural and Positional
                  Node and Graph Embedding},
  booktitleREMOVED    = {Proceedings of the 2021 {SIAM} International Conference on Data Mining,
                  {SDM} 2021, Virtual Event, April 29 - May 1, 2021},
  booktitle    = {{SIAM} Int. Conf. on Data Mining, {SDM} 2021},
  pages        = {163--171},
  publisher    = {{SIAM}},
  address = {USA},
  year         = {2021},
  urlREMOVED         = {https://doi.org/10.1137/1.9781611976700.19},
  doi         = {10.1137/1.9781611976700.19},
  bibsource    = {dblp computer science bibliography, https://dblp.org}
}

@inproceedings{DBLP:conf/cikm/CuiL0Y22,
  author       = {Hejie Cui and
                  Zijie Lu and
                  Pan Li and
                  Carl Yang},
  editor       = {Mohammad Al Hasan and
                  Li Xiong},
  title        = {On Positional and Structural Node Features for Graph Neural Networks
                  on Non-attributed Graphs},
  booktitleREMOVED    = {Proceedings of the 31st {ACM} International Conference on Information
                  {\&} Knowledge Management, Atlanta, GA, USA, October 17-21, 2022},
  booktitle    = {{ACM} CIKM 2022},
  pages        = {3898--3902},
  publisher    = {{ACM}},
  address = {USA},
  year         = {2022},
  urlREMOVED          = {https://doi.org/10.1145/3511808.3557661},
  doi          = {10.1145/3511808.3557661}
}

@article{PhysRevE.78.046110,
  title = {Benchmark graphs for testing community detection algorithms},
  author = {Lancichinetti, Andrea and Fortunato, Santo and Radicchi, Filippo},
  journal = {Phys. Rev. E},
  volume = {78},
  issue = {4},
  pages = {046110},
  numpages = {5},
  year = {2008},
  month = {Oct},
  publisher = {American Physical Society},
  doi = {10.1103/PhysRevE.78.046110},
  urlREMOVED = {https://link.aps.org/doi/10.1103/PhysRevE.78.046110}
}

@inproceedings{chakraborty2014permanence,
  title={On the permanence of vertices in network communities},
  author={Chakraborty, Tanmoy and Srinivasan, Sriram and Ganguly, Niloy and Mukherjee, Animesh and Bhowmick, Sanjukta},
  booktitle={Proceedings of the 20th ACM SIGKDD international conference on Knowledge discovery and data mining},
  pages={1396--1405},
  doi          = {10.1145/2623330.2623707},
  year={2014}
}

@article{al2023community,
  title={Community detection in networks through a deep robust auto-encoder nonnegative matrix factorization},
  author={Al-sharoa, Esraa and Rahahleh, Baraa},
  journal={Engineering Applications of Artificial Intelligence},
  volume={118},
  pages={105657},
  year={2023},
  publisher={Elsevier},
  doi= {10.1016/J.ENGAPPAI.2022.105657},
}

@article{karypis1998fast,
  title={A fast and high quality multilevel scheme for partitioning irregular graphs},
  author={Karypis, George and Kumar, Vipin},
  journal={SIAM Journal on scientific Computing},
  volume={20},
  number={1},
  pages={359--392},
  doi          = {10.1137/S1064827595287997},
  year={1998},
  publisher={SIAM}
}

@inproceedings{kmeans,
  title={Some methods for classification and analysis of multivariate observations},
  author={MacQueen, James and others},
  booktitle={Proceedings of the fifth Berkeley symposium on mathematical statistics and probability},
  volume={1},
  number={14},
  pages={281--297},
  year={1967},
  organization={Oakland, CA, USA}
}

@inproceedings{DBLP:conf/kdd/WangCF13,
author = {Wang, Jia and Cheng, James and Fu, Ada Wai-Chee},
title = {Redundancy-aware maximal cliques},
year = {2013},
isbn = {9781450321747},
publisher = {ACM},
address = {USA},
booktitle = {Proceedings of the 19th ACM SIGKDD International Conference on Knowledge Discovery and Data Mining},
pages = {122–130},
numpages = {9},
location = {Chicago, Illinois, USA},
series = {KDD'13}
}

@inproceedings{DBLP:conf/icde/Li0CZL0021,
  author       = {Xiaofan Li and
                  Rui Zhou and
                  Lu Chen and
                  Yong Zhang and
                  Chengfei Liu and
                  Qiang He and
                  Yun Yang},
  title        = {Finding a Summary for All Maximal Cliques},
  booktitleREMOVED    = {37th {IEEE} International Conference on Data Engineering, {ICDE} 2021,
                  Chania, Greece, April 19-22, 2021},
  pages        = {1344--1355},
  publisher    = {IEEE},
  address = {Los Alamitos, CA, USA},
  year         = {2021},
  urlREMOVED          = {https://doi.org/10.1109/ICDE51399.2021.00120},
  doi          = {10.1109/ICDE51399.2021.00120},
  bibsource    = {dblp computer science bibliography, https://dblp.org}
}

@inproceedings{d2023clique,
  title={Clique-TF-IDF: a new partitioning framework based on dense substructures},
  author={D’Elia, Marco and Finocchi, Irene and Patrignani, Maurizio},
  booktitle={International Conference of the Italian Association for Artificial Intelligence},
  pages={396--410},
  year={2023},
  doi          = {10.1007/978-3-031-47546-7\_27},
  organization={Springer}
}

@article{DBLP:journals/aiopen/LiuT21,
  author       = {Xueyi Liu and
                  Jie Tang},
  title        = {Network representation learning: {A} macro and micro view},
  journal      = {{AI} Open},
  volume       = {2},
  pages        = {43--64},
  year         = {2021},
  urlREMOVED          = {https://doi.org/10.1016/j.aiopen.2021.02.001},
  doi          = {10.1016/J.AIOPEN.2021.02.001},
  bibsource    = {dblp computer science bibliography, https://dblp.org}
}

@article{DBLP:journals/aiopen/ZhaoCCZT22,
  author       = {Shu Zhao and
                  Jialin Chen and
                  Jie Chen and
                  Yanping Zhang and
                  Jie Tang},
  title        = {Hierarchical label with imbalance and attributed network structure
                  fusion for network embedding},
  journal      = {{AI} Open},
  volume       = {3},
  pages        = {91--100},
  year         = {2022},
  urlREMOVED          = {https://doi.org/10.1016/j.aiopen.2022.07.002},
  doi          = {10.1016/J.AIOPEN.2022.07.002},
  bibsource    = {dblp computer science bibliography, https://dblp.org}
}

@inproceedings{networkrepository,
      title = {The Network Data Repository with Interactive Graph Analytics and Visualization},
      author={Ryan A. Rossi and Nesreen K. Ahmed},
      booktitle = {AAAI},
      url ={https://networkrepository.com},
      year={2015}
}

@inproceedings{KONECT,
  author       = {J{\'{e}}r{\^{o}}me Kunegis},
  editorOPTIONAL       = {Leslie Carr and
                  Alberto H. F. Laender and
                  Bernadette Farias L{\'{o}}scio and
                  Irwin King and
                  Marcus Fontoura and
                  Denny Vrandecic and
                  Lora Aroyo and
                  Jos{\'{e}} Palazzo M. de Oliveira and
                  Fernanda Lima and
                  Erik Wilde},
  title        = {{KONECT:} the Koblenz network collection},
  booktitle    = {22nd International World Wide Web Conference, {WWW} '13, Rio de Janeiro,
                  Brazil, May 13-17, 2013, Companion Volume},
  pages        = {1343--1350},
  publisher    = {{ACM}},
  year         = {2013},
  urlREMOVED          = {https://doi.org/10.1145/2487788.2488173},
  doi          = {10.1145/2487788.2488173},
  bibsource    = {dblp computer science bibliography, https://dblp.org}
}

\end{document}